\documentclass[conference]{IEEEtran}
\IEEEoverridecommandlockouts
\usepackage{mathptmx} 

\usepackage{fancyhdr}
\usepackage[normalem]{ulem}
\usepackage[hyphens]{url}
\usepackage[sort,nocompress]{cite}
\usepackage[final]{microtype}
\usepackage[bookmarks=true,breaklinks=true,letterpaper=true,colorlinks,citecolor=blue,linkcolor=blue,urlcolor=blue]{hyperref}
\usepackage{pifont}
\usepackage{comment}
\usepackage{xspace}
\usepackage{hyperref}
\usepackage{makecell}
\usepackage{subcaption}
\usepackage{pifont}
\usepackage{multirow}
\usepackage{algorithm,algorithmic}
\usepackage{arydshln}
\usepackage{cite}
\usepackage{amsmath,amssymb,amsfonts}
\usepackage{graphicx}
\usepackage{textcomp}
\usepackage{xcolor,soul}
\usepackage[hyphens]{url}
\usepackage{tikz}
\usepackage{array}
\newcommand*\circled[1]{\tikz[baseline=(char.base)]{
            \node[shape=circle,draw,inner sep=0.5pt] (char) {#1};}}
\usepackage[export]{adjustbox}
\usepackage{arydshln}
\usepackage{subcaption}
\newcommand{\insertFigure}[2]{
    \begin{figure}[t!]
        \centering
        \includegraphics[width=\linewidth]{fig/#1.pdf}
	\vspace{-6mm}
        \caption{ #2}
	\vspace{-5mm}
        \label{fig:#1}
    \end{figure}
}

\newcommand{\squishlist}{
 \begin{list}{$\bullet$}
  { \setlength{\itemsep}{1pt}
     \setlength{\parsep}{0pt}
     \setlength{\topsep}{3pt}
     \setlength{\partopsep}{0pt}
     \setlength{\leftmargin}{0pt}
     \setlength{\labelwidth}{0pt}
     \setlength{\labelsep}{1pt} } }

\newcommand{\squishnums}{
 \begin{list}{$\bullets$}
  { \setlength{\itemsep}{0pt}
     \setlength{\parsep}{3pt}
     \setlength{\topsep}{3pt}
     \setlength{\partopsep}{0pt}
     \setlength{\leftmargin}{1.5em}
     \setlength{\labelwidth}{1em}
     \setlength{\labelsep}{0.5em} } }

\newcommand{\squishlisttwo}{
 \begin{list}{$\bullet$}
  { \setlength{\itemsep}{0pt}
     \setlength{\parsep}{0pt}
    \setlength{\topsep}{0pt}
    \setlength{\partopsep}{0pt}
    \setlength{\leftmargin}{2em}
    \setlength{\labelwidth}{1em}
    \setlength{\labelsep}{0.5em} } }

\newcommand{\squishend}{
  \end{list}  }

\newcommand{\squishnobullet}{
 \begin{list}{}
  { \setlength{\itemsep}{0pt}
     \setlength{\parsep}{0pt}
     \setlength{\topsep}{3pt}
     \setlength{\partopsep}{0pt}
     \setlength{\leftmargin}{0.5em}
     \setlength{\labelwidth}{1em}
     \setlength{\labelsep}{0.5em} } }
\newcommand{\Accel}{{\textit{FEATHER}}\xspace}
\newcommand{\rir}{{\textit{RIR}}\xspace}
\newcommand{\noc}{{\textit{BIRRD}}\xspace}
\newcommand{\nest}{{\textit{NEST}}\xspace}

\newcommand{\cmark}{\ding{51}}%
\newcommand{\xmark}{\ding{55}}%

\newcommand{\secref}[1]{\S\ref{#1}}
\newcommand{\figref}[1]{Fig.~\ref{#1}}
\newcommand{\tabref}[1]{Tab.~\ref{#1}}

\newcolumntype{L}[1]{>{\raggedright\let\newline\\\arraybackslash\hspace{0pt}}m{#1}}
\newcolumntype{C}[1]{>{\centering\let\newline\\\arraybackslash\hspace{0pt}}m{#1}}
\newcolumntype{R}[1]{>{\raggedleft\let\newline\\\arraybackslash\hspace{0pt}}m{#1}}

\definecolor{deepblue}{rgb}{0,0,0.5}
\definecolor{deepred}{rgb}{0.6,0,0}
\definecolor{deepgreen}{rgb}{0,0.5,0}

\usepackage{listings}
\DeclareFixedFont{\ttb}{T1}{txtt}{bx}{n}{9} 
\DeclareFixedFont{\ttm}{T1}{txtt}{m}{n}{9}  

\newcommand\pythonstyle{\lstset{
language=Python,
basicstyle=\ttm,
morekeywords={self,create,activate,install,clone,load,image},              
keywordstyle=\ttb\color{deepblue},
emph={MyClass,__init__, pip3, python, conda,git, sudo,apt-get},          
emphstyle=\ttb\color{deepred},    
stringstyle=\color{deepgreen},
frame=tb,                         
showstringspaces=false
}}

\lstnewenvironment{python}[1][]
{
\pythonstyle
\lstset{#1}
}
{}

\newcommand\pythoninline[1]{{\pythonstyle\lstinline!#1!}}

\usepackage{cite}
\usepackage{amsmath,amssymb,amsfonts}
\usepackage{algorithmic}
\usepackage{graphicx}
\usepackage{textcomp}
\usepackage{xcolor}
\def\BibTeX{{\rm B\kern-.05em{\sc i\kern-.025em b}\kern-.08em
    T\kern-.1667em\lower.7ex\hbox{E}\kern-.125emX}}

\newif\ifcommenton
\commentontrue

\ifcommenton
\newcommand{\TODO}[1]{\textcolor{red}{[TODO] #1}}
\newcommand{\JT}[1]{{\color{brown}\bfseries [Jianming: #1]}}
\newcommand{\AI}[1]{{\color{blue}\bfseries [Anirudh: #1]}}
\newcommand{\PC}[1]{{\color{blue}\bfseries [Prasanth: #1]}}
\newcommand{\TK}[1]{{\color{violet}\bfseries [TK: #1]}}
\newcommand{\GJ}[1]{{\color{blue}\bfseries [GJ: #1]}}
\newcommand{\fixme}[1]{{{\color{blue} #1}}}
\else
\newcommand{\TODO}[1]{}
\newcommand{\AI}[1]{}
\newcommand{\PC}[1]{}
\newcommand{\JT}[1]{}
\newcommand{\TK}[1]{}
\newcommand{\GJ}[1]{}
\newcommand{\fixme}[1]{}
\fi

\definecolor{colorRevA}{rgb}{0.85, 0.90, 0.98}
\definecolor{colorRevB}{rgb}{0.97, 0.80, 0.796}
\definecolor{colorRevC}{rgb}{0.87, 0.83, 0.90}
\definecolor{colorRevD}{rgb}{0.83, 0.91, 0.83}
\definecolor{colorRevE}{rgb}{1, 0.90, 0.80}
\definecolor{blond}{rgb}{0.98, 0.94, 0.75}

\newif\ifrevisionon
\revisionontrue

\begin{document}

\title{\huge FEATHER: A Reconfigurable Accelerator with Data Reordering Support for Low-Cost On-Chip Dataflow Switching
}

\author{\IEEEauthorblockN{Jianming Tong}
\IEEEauthorblockA{
\textit{Georgia Institute of Technology}\\
Atlanta, Georgia, USA \\
jianming.tong@gatech.edu}
\and
\IEEEauthorblockN{Anirudh Itagi}
\IEEEauthorblockA{
\textit{Georgia Institute of Technology}\\
Atlanta, Georgia, USA \\
aitagi7@gatech.edu}
\and
\IEEEauthorblockN{Prasanth Chatarasi}
\IEEEauthorblockA{\textit{IBM Research} \\
Yorktown Heights, USA \\
prasanth@ibm.com}
\and
\IEEEauthorblockN{Tushar Krishna}
\IEEEauthorblockA{
\textit{Georgia Institute of Technology}\\
Atlanta, Georgia, USA \\
tushar@ece.gatech.edu}
}

\maketitle

\begin{abstract}
The inference of ML models composed of diverse structures, types, and sizes boils down to the execution of different dataflows (i.e. different tiling, ordering, parallelism, and shapes). Using the optimal dataflow for every layer of workload can reduce latency by up to two orders of magnitude over a suboptimal dataflow. Unfortunately, reconfiguring hardware for different dataflows involves on-chip data layout reordering and datapath reconfigurations, leading to non-trivial overhead that hinders ML accelerators from exploiting different dataflows, resulting in suboptimal performance. 
To address this challenge, we propose \Accel, an innovative accelerator that leverages a novel spatial array termed \nest and a novel multi-stage reduction network called \noc for performing flexible data reduction with layout reordering under the hood, enabling seamless switching between optimal dataflows with negligible latency and resources overhead.
For systematically evaluating the performance interaction between dataflows and layouts, we enhance Timeloop, a state-of-the-art dataflow cost modeling and search framework, with layout assessment capabilities, and term it as Layoutloop. We model \Accel into Layoutloop and also deploy \Accel end-to-end on the edge ZCU104 FPGA. \Accel delivers $1.27\sim2.89\times$ inference latency speedup and $1.3\sim6.43\times$ energy efficiency improvement compared to various SoTAs like NVDLA, SIGMA and Eyeriss under ResNet-50 and MobiletNet-V3 in Layoutloop. On practical FPGA devices, \Accel achieves $2.65$/$3.91\times$ higher throughput than Xilinx DPU/Gemmini. 
Remarkably, such performance and energy efficiency enhancements come at only $6$\% area over a fixed-dataflow Eyeriss-like accelerator. Our code is released at \url{https://github.com/maeri-project/FEATHER}.
\end{abstract}



\section{Introduction}
\label{sec:introduction}

The field of Machine Learning (ML), specifically Deep Neural Networks (DNNs) is pervasive today across image classification~\cite{image_segmentation,image-classification}, object detection~\cite{object_recognition,pedestrian_detection}, text summarization~\cite{text_summarization} and sentiment analysis~\cite{sentiment_analysis}. Such a plethora of ML models introduces great diversity in structure (serial or parallel layers connectivity), layer types (depth-width, point-width, dilation convolutions, or even a fusion of them), and sizes (number of channels, kernels, height, and width)~\cite{ML_Model_survey_2022, behnam2023subgraph}.

\insertFigure{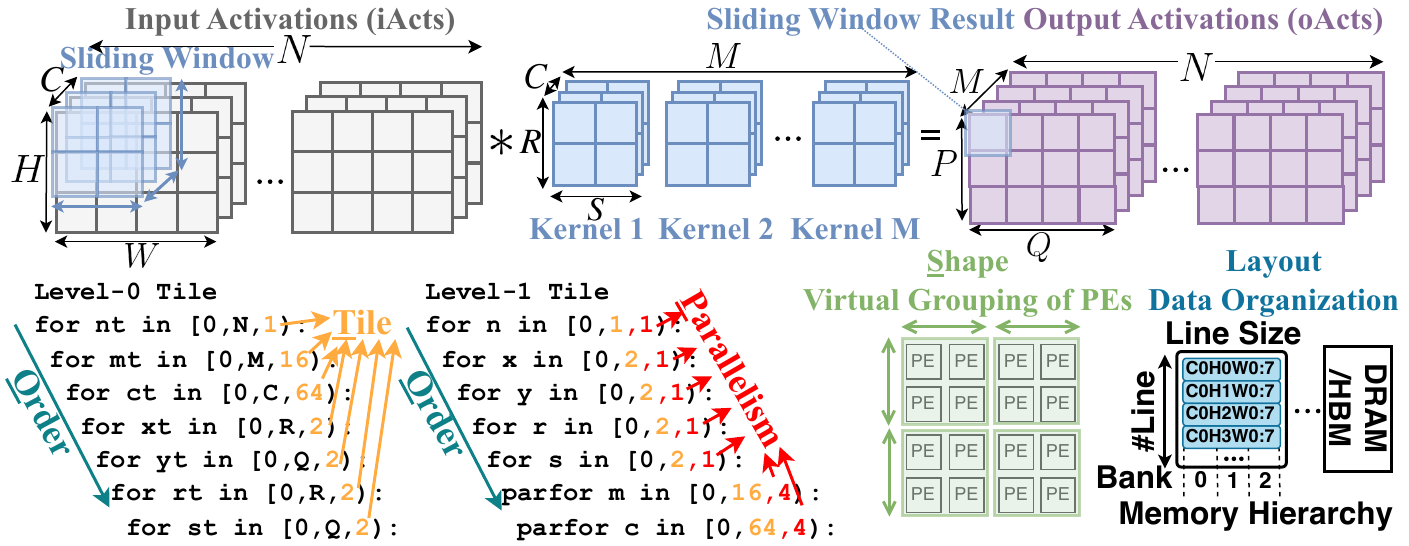}{Terminology of convolution workload and dataflow}

The mechanism for orchestrating a DNN layer 
over the accelerator's on-chip compute and memory resources is called \textit{dataflow}. It can be precisely defined by transformations of the loop nest, as shown in \figref{fig:workload_dataflow}.
Several prior works~\cite{parashar2019timeloop,maestro} have demonstrated that dataflows can lead to significant differences in compute utilization and up to two orders of magnitude variance in latency and energy, and thereby motivated the need to support per-layer dataflow flexibility.

Changing dataflows on accelerators requires (a) reconfiguring datapaths in computation, distribution, and reduction networks, and (b) modifying data layout in on-chip buffers. 
Almost all prior works have focused on the first aspect, and several clever interconnect topologies for data distribution and reduction have been proposed that activate subset of paths at runtime through reconfiguration depending on the dataflow being run~\cite{SARA,sigma}.
However, data layout in the on-chip buffer is a critical and often overlooked in past work. 

\begin{figure}[t!]
    \centering
    \includegraphics[width=\linewidth]{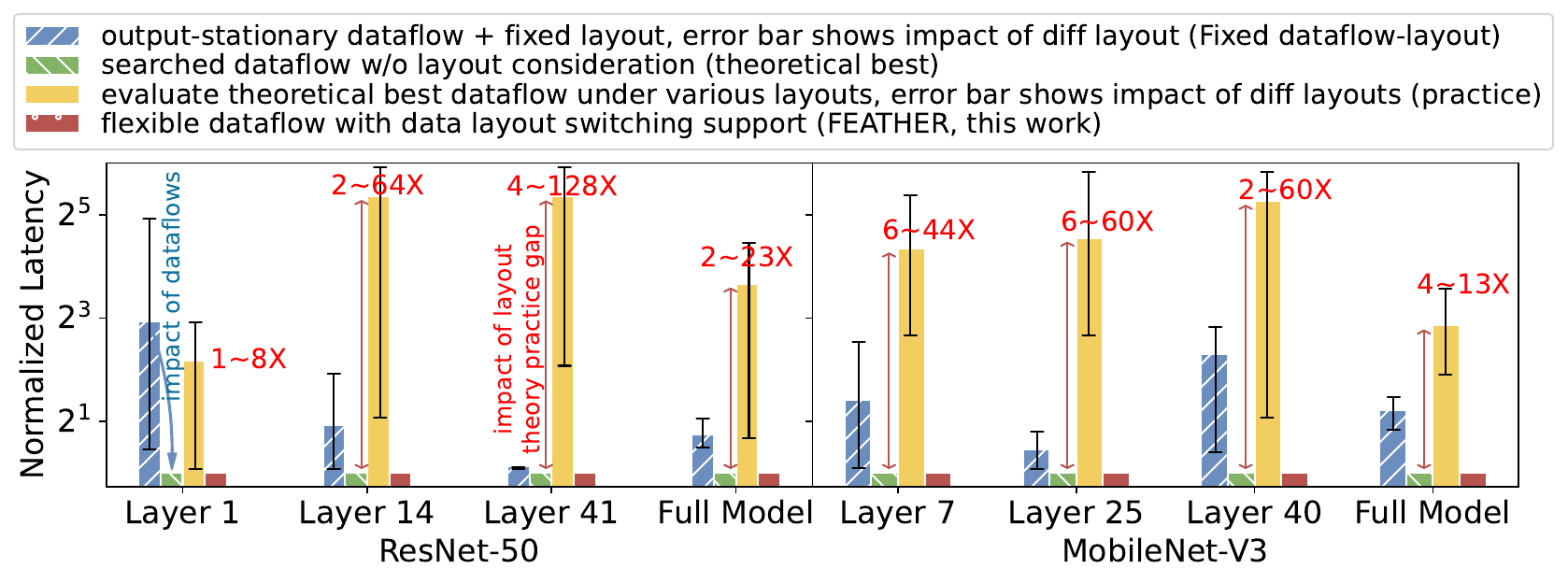}
    \caption{Latency evaluation of dataflows on $16\times 16$ PE array with various layouts (error bar shows layout impacts, less latency is better). The best flexible dataflow (green bar) \textit{theoretically} reduces overall latency of fixed dataflow-layout (blue bar) by $63.3$\%. However, ignoring the impact of layout considerations in theoretical dataflows results in up to a $128\times$ latency gap in \textit{practice} (yellow bar). 
    FEATHER eliminates the gap by co-switching dataflow-layout (red bar).} 
    \label{fig:per_layer_latency}
    \vspace{-2mm}
\end{figure}

In this work, we demonstrate that the high performance of dataflows is unachievable in practice without layout reordering capability. This is because, without a suitable data layout, the required data may be located in the same SRAM banks and compete at the same SRAM reading ports. Such bank conflict slows down the delivery of data to computation engines, leading to stalling and computation underutilization. Overlooking layout reordering thus introduces a significant $128\times$ performance gap between theory and practice as quantified in \figref{fig:per_layer_latency}. We discuss this with more depth in \secref{sec:motivation}.

\begin{table*}[!tp]
    \centering
    \scriptsize
    \caption{Feature comparison: how FEATHER resolves challenges of prior works without on-chip layout reordering.}
    \label{tab:cmp_other_accel}
    \begin{tabular}{ccccccc}
        \hline
            Work & Dataflow Switching  & Layout Reorder  & Challenge & \textbf{\Accel} solution (key component) \\
            \hline
            NVDLA~\cite{nvdla} &\xmark  & no reorder &  underutilization from fixed parallelism & flexible dataflows (\nest) \\
            Xilinx DPU~\cite{xilinx_dpu}, Gemmini~\cite{gemmini} &\xmark & no reorder & linear reduction& parallel logarithmic reduction (\noc) \\
            SIMBA~\cite{simba}, Eyeriss~\cite{chen2016eyeriss} &\xmark & no reorder & load imbalance across PE & pick load-balance dataflows (\nest) \\
            Eyeriss\_v2~\cite{chen2018eyeriss}, SARA~\cite{SARA}  &\cmark & off-chip  & high latency of moving data off-chip & on-chip reordering with latency hidden (\noc) \\ 
            MAERI~\cite{kwon2018maeri}, SIGMA~\cite{sigma} &\cmark & off-chip & long wires of reduction network & small standalone reduction network (\noc) \\
        \hline
    \end{tabular}
    \vspace{-4mm}
\end{table*}

Unfortunately, layout reordering comes with severe latency and energy overheads. Off-chip layout reordering requires back-and-forth data movement between off-chip DRAM/HBM and computation, while on-chip layout reordering requires additional intermediate storage and extra latency in the critical path. In fact, these costs can outweigh the benefits of switching dataflows, leading existing ML accelerators to compromise settling on a single dataflow (e.g., Xilinx DPU, Gemmini, NVDLA, Eyeriss in Table~\ref{tab:cmp_other_accel}) that provides good average utilization across all layers, but sub-optimal performance.

To unleash optimal performance, we propose a novel accelerator \Accel, \underline{F}lexible \underline{E}ngine for \underline{A}cceleration of \underline{T}ensors with \underline{H}ardware \underline{E}lement for \underline{R}eordering, which includes a novel reconfigurable reduction network called \underline{B}utterfly \underline{I}nterconnect for \underline{R}eduction and \underline{R}eordering in \underline{D}ataflows (\noc). 
With \noc, the latency of layout reordering is completely hidden in data reduction, allowing data layout in on-chip storage to be manipulated for the demand of optimal dataflow without any latency costs. 
We call this approach as \underline{r}eordering \underline{i}n data \underline{r}eduction (RIR).
Thus, \Accel fully achieves the theoretical performance of optimal dataflows without incurring bank conflicts.
Furthermore, \Accel also pioneers a new paradigm to co-switch both dataflows and data layouts at layer granularity, with minimal switching overheads. This ability to accommodate low-cost layout-dataflow co-switching is, as far as we know, unsupported by any existing accelerator.

To fully explore the potential of \Accel, we also developed a tool that facilitates: (a) dataflow evaluation factoring in data layout, and (b) (layout, dataflow) co-exploration.

\noindent Our key contributions can be summarized as follows:

\squishlist
\item We demonstrate the interaction between dataflows and data layouts, motivating the need for data reordering support within reconfigurable dataflow accelerators. 
We further categorize existing reordering patterns and implementations (\secref{sec:motivation}). 
\item We present a novel accelerator \Accel with several novel features (\secref{sec:system_overview}). First, a neural engine with temporal local reduction and spatial forwarding, \nest, for dataflow flexibility. Second, a multi-stage network called \noc enabling flexible reductions from arbitrary groups of multiple inputs to multiple results, at lower area overhead compared to prior works with similar capabilities. Further, BIRRD supports Arbitrary reorder via a novel technique RIR, that completely conceals data layout reordering latency behind reduction (\secref{sec:rir}).
\item We extend a state-of-the-art accelerator modeling framework Timeloop~\cite{parashar2019timeloop} with support for physical on-chip storage, layout representation, and dataflow-layout co-search. We call this new framework \textit{Layoutloop} (\secref{sec:layoutloop}) and use it for our evaluations.
\item We implement and deploy \Accel, end-to-end, on an edge ZCU 104 FPGA device and also model it using \textit{Layoutloop}. \Accel achieves $1.27\sim2.89\times$ inference latency speedup and $1.3\sim6.43\times$ energy efficiency improvement compared to various SoTAs across multiple DNN models, and $2.65$$\times$/$3.91\times$ more throughput than Xilinx DPU/Gemmini on real FPGAs. On average, efficient pairs of (dataflow, layout) results in an energy savings of 27\% to 33\% across workloads despite the energy costs of layout reordering. Remarkably, all enhancements come at only $6$\% area over a fixed-dataflow Eyeriss-like accelerator.
\squishend

\section{Background and Motivation}
\label{sec:motivation}

\subsection{Dataflow Space in Convolution}
\label{sec:dataflow_terminology}

\figref{fig:workload_dataflow} depicts a convolution operation with seven dimensions with various shapes. Dataflows can be represented as a nested loop with four types of optimizations~\cite{Flexion, Union}.

\begin{table}[!t]\centering
\caption{On-chip memory terminology}\label{tab:terminology}
\vspace{-1mm}
\scriptsize
\begin{tabular}{cc}\hline
Term & Meaning \\ \hline
Buffer & \makecell{A logical 2D on-chip memory (num\_line $\times$ line\_size) stacking multiple \\ 
SRAM banks both vertically (num\_line) and horizontally (line\_size).} \\
Bank & A physical 2D SRAM (entries $\times$ io) with address/data ports. \\
Line/Row & A buffer line (line\_size = accumulated IO of horizontal SRAM banks). \\
Port & An input/output port, each bank has at most two ports in TSMC 28nm. \\
\hline
\end{tabular}
\vspace{-4mm}
\end{table}

\label{sec:tops}
\squishlist
\item \textbf{\underline{(T)}iling} breaks down dimensions of iActs $N, C, H, W$ into smaller chunks, and enables executing workloads in tile granularity as on-chip storage is limited. 
\item \textbf{\underline{(O)}rdering} allows arbitrary loop reordering (aka ``stationarity"~\cite{chen2016eyeriss}) to reuse more data since dimensions $N,M,C,P,Q,R,S$ do not come with loop-carried dependencies except reduction-dependencies over C, R, and S.
\item \textbf{\underline{(P)}arallelism} allows for arbitrary parallelism over any dimensions as all dependencies are loop-independent, leading to different spatial reuse opportunities. 
\item \textbf{\underline{(S)}hape} defines the virtual grouping of the physical PE array.
\squishend

These \textit{dataflow flexibility (TOPS)}~\cite{Flexion} create an extremely large dataflow design space with a complexity of $O(10^{36})$ for a single convolution layer~\cite{gamma}. 
The choice of the dataflow affects both runtime performance (as it affects overall compute utilization) and energy efficiency (as it affects the number of accesses across the memory hierarchy).
Not surprisingly, no single dataflow is generally optimal for all types of layers given their diverse sizes and shapes~\cite{maestro,parashar2019timeloop}. This can be seen by comparing the first two bars (blue and green bars) in \figref{fig:per_layer_latency}.

\subsection{Data Layout in on-chip Storage}
Various organizations of on-chip storage are logically a 2D buffer (\tabref{tab:terminology}), where the width of each logical buffer row, termed ``line size", represents bandwidth (max number of data words a buffer could supply per cycle) and the depth represents the total number of buffer row entries as shown in \figref{fig:workload_dataflow}. 

Physically, on-chip storage is implemented by BRAM/URAM in FPGA and SRAM in ASICs, which come with a \textit{fixed number (often two) read or write ports}. Therefore, once arranged into the logical 2D buffer, the number of lines being concurrently accessed is limited by the number of ports. A request that accesses more lines than the available ports will lead to bank conflicts, resulting in a slowdown from the reading/writing delay (resource hazard).

\begin{figure}[t!]
    \centering
    \includegraphics[width=\linewidth]{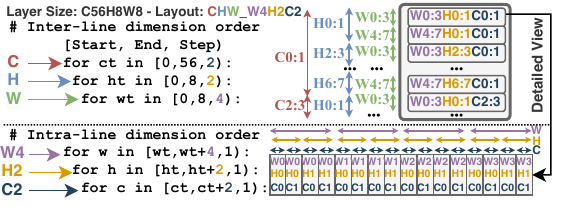}
    \caption{Layout terminology example: `CHW\_W4H2C2'. `CHW' signifies the inter-line dimension order as C$\rightarrow$H$\rightarrow$W across lines. `W4H2C2' indicates the intra-line dimension order: (4,2,2) elements from the (W,H,C) dimensions are flattened into a single row in the order of W$\rightarrow$H$\rightarrow$C.}
    \label{fig:NTLR_Layout_example_simple}
    \vspace{-2mm}
\end{figure}

\textbf{Data Layout Terminology.} In this paper, data layout is represented as ``(Inter-line dimension order)\_(Intra-line dimension order interleaved with sizes)" with one example shown in \figref{fig:NTLR_Layout_example_simple}. For instance, two commonly used PyTorch data layouts, channel-last~\cite{channel_last_layout} and row-major~\cite{row_major_layout}, can be interpreted as Channel (C) or Width (W) being the innermost dimension in both inter and intra-line orders, separately.

\subsection{Interaction of Dataflow and Data Layout}

\begin{figure*}[ht!]
    \centering
    \includegraphics[width=0.9\textwidth]{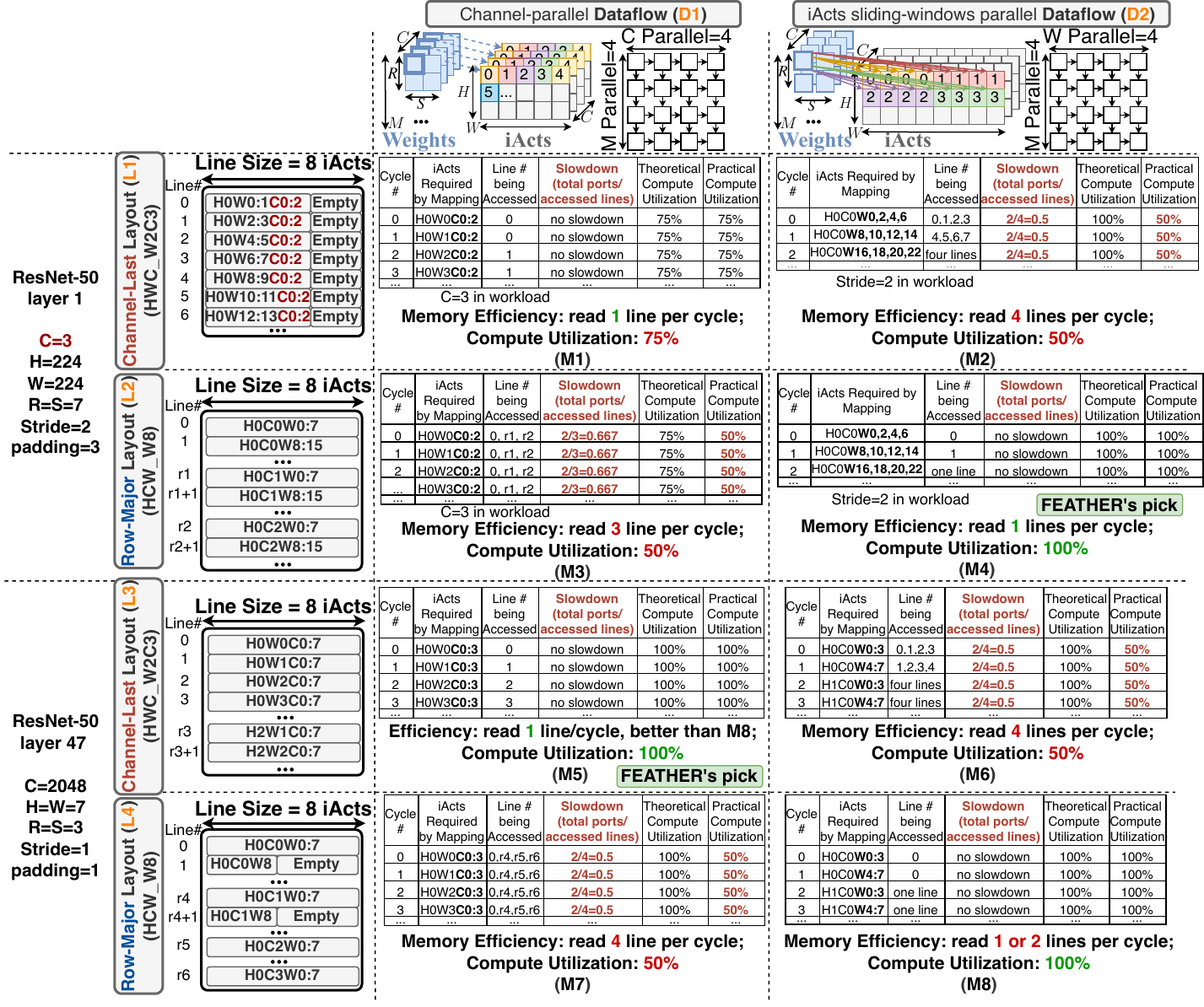}
    \caption{Memory efficiency and computation utilization of various \textbf{(workload, dataflow, data layout)} combinations on weight-stationary $4\times4$ Systolic Array (SA). \textbf{Dataflows:} input channel-parallel (D1) and sliding-window parallel (D2). Dataflow D1/D2 reads at most four iActs from C/W dimension concurrently from the on-chip buffer every cycle, separately. The digit in iActs indicates the cycle index such iActs get read. \textbf{Workloads:} (1) ResNet-50 layer 1 with a large height and width, and (2) ResNet-50 layer 47 with a large channel number. \textbf{Layouts:} channel last-layout (L1, L3) and row-major layout (L2, L4).
In the channel-last layout, data from different input channels (dimension $C$) are spread across an individual line, while in the row-major layout, multiple data from different input width (dimension $W$) are flattened. 
The performance of mappings (M1$\sim$M8) for different (workload, dataflow, layout) combinations are analyzed in the tables.
In each table, ``iActs Required by Mapping" lists all iActs that need to be concurrently read from on-chip buffer every cycle, and the corresponding index (\#) of lines being accessed are listed in ``Line \# being Accessed". We assume dual read ports (because TSMC offers SRAM with at most two ports), such that a concurrent read for more than two lines leads to slowdown, which reduces ``Theoretical Computation Utilization" (estimated as mapping efficiency over the array) into ``Practical Compute Utilization" (computed as multiplication of theoretical utilization with slow down). \textbf{Takeaway:} For optimal performance, co-switching (dataflow, layout) is crucial, because \textit{dataflow} matters (comparing M1 vs. M4), and \textit{layout} also matters (comparing M2 vs. M4).}
\label{fig:Motivation}
\end{figure*}

In the rest of the paper, we refer to a (dataflow, layout) pair with bank conflicts as \textit{\textbf{discordant}}, whereas its non-conflicting counterpart is termed \textit{\textbf{concordant}}, i.e. a layout is concordant to a dataflow if there are no bank conflicts. 
And we use \textit{\textbf{concordant dataflow space of a layout}} to refer to all concordant dataflows choices under a layout.
Switching optimal dataflows for different layers is not trivial given that it necessitates a costly reordering to convert the data layout into a concordant form to prevent bank conflicts. 

In this subsection, we discuss some crucial insights, underscoring the necessity of co-switching dataflows-layouts for different layers by evaluating the performance of various combinations of dataflows and data layouts as shown in \figref{fig:Motivation}. 

\noindent {\textit{Insight 1: Discordance between dataflow and data layout leads to bank conflicts and results in performance degradation.}}

A discordance between dataflow and data layout leads to slowdown because compute units have to stall and wait for data to arrive, as illustrated by the slowdown from green bar to yellow bar in \figref{fig:per_layer_latency}. Taking ResNet-50 layer 47 as an example (\figref{fig:Motivation}-M7), the channel-parallel dataflow requires concurrent access to iActs (H0W0C0:3), which are distributed across four separate lines, including line 0, r4, r5 and r6, in the row-major layout (\figref{fig:Motivation}-L4). Therefore, a $0.5$ slowdown is encountered, resulting in $50\%$ practical computation utilization. Such bank conflicts cannot be resolved by line rotation, since moving one conflicted line to another bank leaves the remaining three lines still in conflict. This slowdown analysis also applies to \figref{fig:Motivation}-M1,2,3,6. 

\noindent {\textit{Insight 2: Co-switching (dataflow, layout) for different layers is necessary for high performance with optimal efficiency.}}\hfill

For certain workloads, picking a fixed layout might not suffer a slowdown from bank conflicts, like choosing row-major layout for both two layers of ResNet-50 (M4 and M8 in \figref{fig:Motivation}). However, the mapping M5 (``\Accel's pick") delivers better energy efficiency than M8 as it supplies data with reading less number of lines. 
Therefore, even under a small parallelism of four, co-switching dataflows and layouts is essential to maximize performance and energy efficiency. Practical designs (e.g. $128\times 128$ systolic array in Google TPU) will further amplify such a need as it brings higher parallelism in more dimensions and requires more concurrent data.

\noindent {\textit{Insight 3: Systematic layout modeling should be factored into dataflow exploration for bridging the theory-practice gap.}}

Dataflow has a huge space, which requires systematic modeling and searching algorithms to identify the optimum. 
However, many dataflow exploration frameworks~\cite{maestro, parashar2019timeloop} and algorithms~\cite{gamma,digamma,magma,Marvel} purely model on-chip storage as bandwidth, often assuming ideal data layouts, which could lead to significant theory-practice performance gap. 
For instance, all layouts in \figref{fig:Motivation} possess identical bandwidth, but they result in markedly different compute utilization and energy efficiency for two workloads, which is not the case in the existing frameworks as they do not model layout.
In \figref{fig:per_layer_latency}, we find that the best dataflow reported by a mapper from an existing framework~\cite{parashar2019timeloop} (green bar), can in practice perform 2 orders of magnitude worse (yellow bar) than the fixed dataflow case (blue bar) due to the discordant accesses to the on-chip memory. 
Thus, taking layout into consideration \textit{during} search (red bar) is necessary and crucial.

\begin{figure*}[t]
 \centering
    \subfloat[Initial layout \label{fig:typical_2d_buffer}]{{\includegraphics[width=0.115\linewidth]{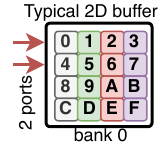}}}
    \subfloat[Line Rotation \label{fig:line_rotation}]{{\includegraphics[width=0.2\linewidth]{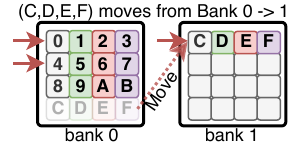}}}
    \subfloat[Transpose \label{fig:transpose}]{{\includegraphics[width=0.10\linewidth]{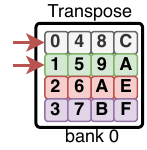}}}
    \subfloat[Row-Reorder \label{fig:row_reorder}]{{\includegraphics[width=0.12\linewidth]{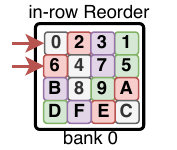}}}
    \subfloat[Arbitrary Reorder \label{fig:arbitrary_reorder}]{{\includegraphics[width=0.17\linewidth]{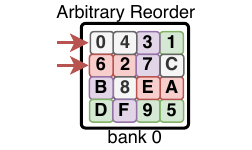}}}
    \subfloat[Concordant Dataflow Space \label{fig:reorder_cmp}]{{\includegraphics[width=0.22\linewidth]{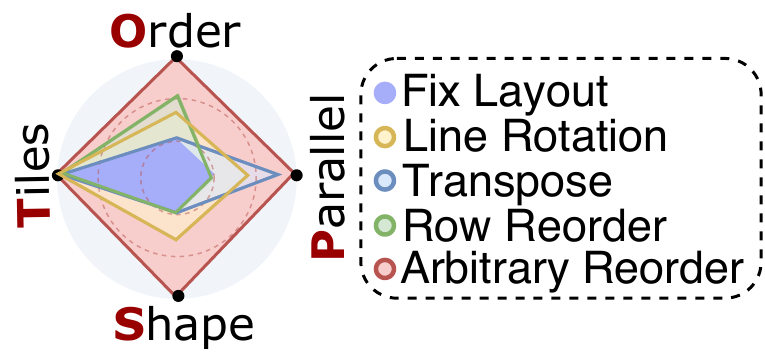}}}
    \vspace{-5mm}
    \caption{Overview of reordering \textit{patterns}. The 2D layout without any reordering is shown in~\ref{fig:typical_2d_buffer}, which only allows reading two rows concurrently, assuming true dual-port SRAM. 
    Line Rotation (\ref{fig:line_rotation}, e.g., Medusa~\cite{Medusa}) moves a row from bank 0 to bank 1 prior to reading, enabling simultaneous access to at most three rows from bank 0 through dual-bank ports. This technique, however, utilizes additional port from bank 1, potentially limiting access to other data in bank 1. 
    Transpose (\ref{fig:transpose}, e.g., MTIA~\cite{MITA} and TPUv4i~\cite{tpuv4i}) could swap rows with columns. Row Reorder (\ref{fig:row_reorder}, e.g., TPUv4i~\cite{tpuv4i}) permutes data within each row. Arbitrary reorder (\ref{fig:arbitrary_reorder}, proposed in this work) enables arbitrary permutation for data within the entire 2D buffer. Line Rotation, Transpose and Row-Reorder are done by prior works by reading at most two rows per bank, leverage Transpose/Permute unit to reorder and then write data back in concordant order (On-chip RAR in~\ref{fig:on_chip_reorder}). In contrast, \Accel's \noc network (\secref{sec:afft}) performs the Arbitrary-Reorder during the reduction phase of the matrix multiplication or convolution computation (RIR in \figref{fig:reorder_in_reduction}). The concordant dataflow space supported by each layout reorder pattern is shown in \ref{fig:reorder_cmp}. \textit{Reordering enables a given layout to alter the order of data it could provide \textbf{per cycle} and \textbf{across cycles}.} Among four dimensions (T,O,P,S) of concordant dataflow space, reordering enlarges O,P,S by supporting dataflows to read from or write to layout in different order. Note that reordering by itself cannot enlarge T dimension flexibility because higher \underline{T}iles flexibility requires accessing more data per cycle.}
    \label{fig:sota_reordering_concept_cmp}
    \vspace{-4mm}
\end{figure*}

    
\begin{figure*}[t]
    \centering
     \subfloat[Off-chip Data Reorder. \label{fig:off_chip_reorder}]{{\includegraphics[width=0.31\textwidth]{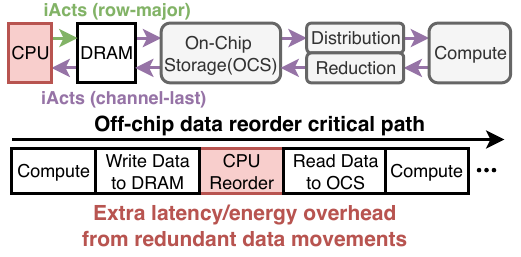}}}
    \subfloat[Reorder after Reduction (prior works). \label{fig:on_chip_reorder}]{\makebox[0.31\textwidth][c]{\includegraphics[width=0.26\textwidth, scale=0.7]{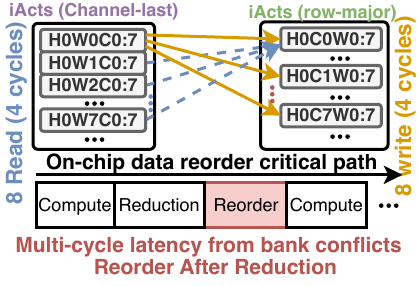}}}
    \subfloat[Reorder in Reduction (RIR, this work). \label{fig:reorder_in_reduction}]{\includegraphics[width=0.3\textwidth]{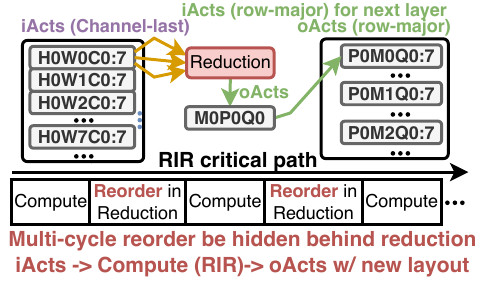}}
    \vspace{-5mm}
    \caption{Comparison of data reordering \textit{implementations}. This work proposes RIR that eliminates reorder latency and bank conflicts. We discuss on-chip reorder patterns, including transpose, line rotation, row-reorder and arbitrary reorder, in~\figref{fig:sota_reordering_concept_cmp}.}
    \vspace{-3mm}
    \label{fig:reorder_overview}
\end{figure*}

\subsection{Data Reordering Patterns}
\label{sec:reordering_patterns}
\subsubsection{Reorder Target (iActs)} As established above, both weights and input activations (iActs) necessitate layout reordering within the on-chip memory when switching dataflows.
For ML inference, the structure and weights of ML models are established prior to deployment, enabling the offline optimal dataflow-layout determination for each layer and offline reordering of all weights. Consequently, an optimal layout for weights within the on-chip scratchpad is assured.
However, iActs are generated in real-time, so that iActs reordering happens online. Therefore, this work focuses on layout reordering of iActs. 

\subsubsection{Reorder Patterns vs. Implementations}
Layout transformations require certain reorder capabilities, referred to as reorder patterns. A reorder pattern has different hardware implementations \textit{with different critical-path latency}. To decouple the concept of reorder patterns from their physical implementations, we analyze reordering in two steps: (1) categorize reordering into distinct functional patterns, as illustrated in \figref{fig:sota_reordering_concept_cmp}, and analyze its impact on dataflow flexibility in \secref{lab:reorder_impact}. (2) pinpoint specific hardware implementations to these patterns in \secref{sec:reorderin_impl_ref}.

\subsubsection{Impact of Reorder Patterns on Dataflow Flexibility}
\label{lab:reorder_impact}
A fixed layout has limited concordant dataflow space, restricting fully-flexible accelerators to less-performant dataflow choices. To improve performance, reordering is required to enlarge concordant dataflow space with more flexibility in TOPS.
\squishlist
\item \textbf{Fixed layout} (\figref{fig:typical_2d_buffer}) is only concordant to dataflows which concurrently access up-to two rows within a single bank, such as $(0,1,2,\cdots,7)$. This restricts concordant dataflow space to limited T,O,P,S flexibility (see purple quadrilateral in \figref{fig:reorder_cmp}).
\item \textbf{Line Rotation} (\figref{fig:line_rotation}) arguments concordant dataflow space to concurrently access up-to \textbf{three} rows within a single bank by storing a copy of a row in other banks.
For example, to access three rows including data $(0,1,\cdots,7,C,D,E,F)$ from bank 0 in \figref{fig:line_rotation}, row ($C,D,E,F$) is moved to bank such that it provides $(0,1,\cdots,7)$ from bank 0 and $(C,D,E,F)$ from bank 1 to avoid bank conflicts. However, line rotation comes at the price of (1) extra bandwidth: it employs three ports for reading data that could be accessed with up-to two ports under concordant layout, (2) storage: it stores a copy of ($C,D,E,F$). Such price could have been used for supporting more parallelism under arbitrary reordering to improve performance.
\item \textbf{Transpose} (\figref{fig:transpose}) enables concurrently access to up-to two rows \textit{or columns} within a bank, hence augmenting concordant dataflow choice with higher P flexibility than fixed layout. But pure transpose falls short of supporting tiled layout transformation, such as changing layout from HWC\_W2C3 (\figref{fig:Motivation}, L1) to HWC\_W8 (\figref{fig:Motivation}, L2)
\item \textbf{Row Reorder} (\figref{fig:row_reorder}) does not support more concurrent access within a single bank, but enables arbitrary order within each row, hence supporting dataflows with higher O flexibility. Further, row reorder also supports im2col~\cite{chellapilla2006high}, which does not reduce bank conflicts because it still accesses the same number of rows from on-chip buffers.
\item \textbf{Arbitrary Reorder} (\figref{fig:arbitrary_reorder}) enables arbitrary layout transformations, hence making all dataflows concordant with full-fledged O,P,S flexibility, as shown by red diamond in \figref{fig:reorder_cmp}.
\squishend

\begin{table}[!t]\centering
\captionof{table}{SoTA on-chip reordering vs. \Accel.}
\label{tab:comparision_reordering}
  {
    \begin{tabular}[b]{cccc}\hline
      Work & Dataflow  & On-chip Reorder Patterns & Implement \\ \hline
       im2col~\cite{chellapilla2006high} & N/A & Row-Reorder (\figref{fig:line_rotation}) & RAR\\
       Medusa~\cite{Medusa} & N/A & Line Rotation (\figref{fig:line_rotation}) & RAR\\
       MTIA~\cite{MITA} & TOP  & Transpose (\figref{fig:transpose}) & RAR\\
      TPUv4~\cite{tpuv4i} & TO  & Trans.+Row-Reorder (\figref{fig:row_reorder}) & RAR\\
      \textbf{This Work} & TOPS & Arbitrary Reorder (\figref{fig:arbitrary_reorder}) & \textbf{RIR}\\ \hline 
    \end{tabular}
    }
    \vspace{-7mm}
\end{table}

\subsection{Data Reordering Implementations}
\label{sec:reorderin_impl_ref}

The layout reorder patterns described in \figref{fig:sota_reordering_concept_cmp} could have different implementations with \textit{different critical-path latency}.
\subsubsection{\textbf{Existing Implementations}} We classify existing reordering implementations into three categories.
\noindent \paragraph{No Reordering} If there is no reordering, either the accelerator needs to run a fixed dataflow or a subset of dataflows that are concordant to the fixed layout, or pay the cost of bank conflicts due to discordant accesses. This can lead to sub-optimal performance (as shown by blue bar in \figref{fig:per_layer_latency}).

\noindent \paragraph{Off-chip Reordering} SoTA that support dataflow switching (\tabref{tab:cmp_other_accel}) require iActs to move to off-chip DRAM, get reordered there by CPU, and then move back to the accelerator. This naturally incurs extra latency and energy costs (\figref{fig:off_chip_reorder}).

\noindent \paragraph{On-chip Reorder After Reduction (\textbf{RAR})}
Existing on-chip reordering techniques essentially perform reordering after reduction. The post-reduction oActs are first written to the on-chip buffer, then read and sent to a separate unit to perform a layout transformation, and then fed back to compute unit as iActs of the next layer. This puts reordering in the critical path, as shown in~\figref{fig:on_chip_reorder}.
Previous arts all fall into this bucket with \textit{explicit reordering latency}, as listed in~\tabref{tab:comparision_reordering}.  
For example, Medusa~\cite{Medusa} proposes dedicated hardware between on-chip buffer to compute unit to implement line rotation (\figref{fig:line_rotation}); Meta's MTIA~\cite{MITA} proposes a Memory Layout Unit (MLU) to implement transpose; Google's TPUv4~\cite{tpuv4i} also supports row-reordering (\figref{fig:row_reorder}) to facilitate im2col.

\subsubsection{\textbf{Proposed Implementation} - On-Chip Reorder In Reduction (RIR)}
\label{sec:reorder_analysis}
This work proposes to perform reordering on output during reduction phase of computation, such that oActs are written in the layout concordant with the dataflow of the next layer. We call this Reorder in Reduction  (\textbf{RIR}). RIR \textit{implicitly} modifies the layout during the reduction process when \textbf{\textit{generating oActs}} instead of transforming iActs from one layout to another, as depicted in~\figref{fig:reorder_in_reduction}. This approach (i) removes reordering from critical path, (ii) reduces the total number of partial sums into fewer final sums, reducing buffer access and effectively minimizing potential bank conflicts. \secref{sec:rir} provides more details. 
    
\subsection{Inefficiency of SoTA Reconfigurable Dataflow Accelerators}

\textbf{Data Reordering Support.} Driven by the observation that on-chip dataflow plays a crucial role (\secref{sec:dataflow_terminology}), there has been a suite of past work on accelerators with hardware support for running diverse dataflows~\cite{krishna2020data}.
Their key observation is that different dataflows trade-off spatial and temporal reuse, and thereby flexible dataflow requires support for different operand stationarity within buffers and variable-sized spatial and temporal reductions through the interconnect.
Unfortunately, these accelerators have \textbf{two} limitations as elaborated in \secref{sec:reordering_patterns} and \secref{sec:reorderin_impl_ref}: (i) either they do not support any on-chip reordering (\tabref{tab:cmp_other_accel}) or support limited transformations including transpose, line rotation or row-reorder (\tabref{tab:comparision_reordering}). This work extends support to arbitrary reordering. (ii) prior on-chip reordering support can cause bank conflicts, increasing reordering time. This work removes reordering from critical path by doing it during the reduction phase of the computation. 

\textbf{Dataflow-Layout Co-Search.} There has also been a suite of dataflow/mapping search tools~\cite{parashar2019timeloop, gamma, cosa} that can recommend the optimal dataflow given a layer and hardware resources. \textit{However, none of these tools explore on-chip data layouts as part of the search process.}

\textbf{ Contributions of this work.} This work addresses the aforementioned gaps via three key contributions: (i) a reconfigurable accelerator \Accel with a novel on-chip fabric called \noc that provides support for \textit{both} dataflow flexibility and layout flexibility through arbitrary reorder, (ii) a new on-chip data reordering mechanism called RIR (implemented by \noc) whose key goal is to \textit{generate} data in the layout required by the next layer instead of explicitly requiring layout conversion (\secref{sec:rir}), (iii) a tool called LayoutLoop for dataflow and layout co-exploration (\secref{sec:layoutloop}).
\Accel provides two specific benefits over prior work in data reordering: (i) supporting arbitrary reorder, and (ii) proposing RIR to hide reordering latency behind computation, and minimize bank conflicts.

\section{\Accel Overview}
\label{sec:system_overview}

\begin{figure}[t!]
    \centering
    \includegraphics[width=0.8\linewidth]{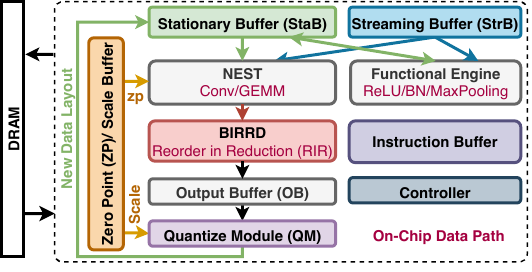}
    \caption{Overview of \Accel architecture. The compute pipeline (NEST$\rightarrow$BIRRD$\rightarrow$OB$\rightarrow$QM) reads iActs from StaB Ping (or Pong) and writes oActs to StaB Pong (or Ping) \textbf{with a new data layout}. }
    \label{fig:lambda_arch}
    \vspace{-3mm}
\end{figure}

\begin{figure}[t!]
    \centering
    \includegraphics[width=0.9\linewidth]{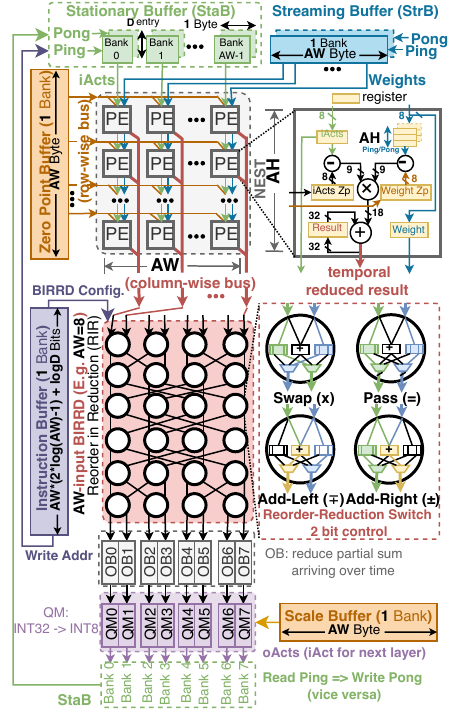}

    \caption{Micro-architecture of \Accel's datapath for convolution/GEMM. For convolution, the NEST reads iActs from StaB and weights from StrB, streaming both in a top-to-bottom pipeline. PEs in a column time-multiplex a common output bus. \noc conducts global spatial reduction and reorders results for targeted StaB banks during reduction, altering data layout in StaB. NEST facilitates inter-layer pipelining by reading iActs from StaB Ping (or Pong) and writes oActs (next-layer iActs) back to StaB Pong (or Ping). Note: \Accel is scalable architecture and we show 8-input \noc as an example.}
    \vspace{-4mm}
    \label{fig:lambda_micro_arch}
\end{figure}

In this section, we provide an overview of \Accel architecture in \figref{fig:lambda_arch} and its micro-architectures in~\figref{fig:lambda_micro_arch}.

\subsection{\Accel's Neural Engine -- NEST}
Accelerators typically use tens of thousands of PEs organized in 1D arrays like MAERI~\cite{kwon2018maeri} and SIGMA~\cite{sigma}, or 2D arrays like Google TPUv4~\cite{tpuv4i} and Meta's MTIA~\cite{MITA}. 
2D PE arrays have better scalability but are limited in their dataflow options due to their rigid structure, leading to suboptimal utilization due to mismatch of layer shapes and array aspect ratios, as prior works have shown~\cite{samajdar2018scale,kwon2018maeri}.
1D arrays with flexible distribution and reduction NoCs~\cite{krishna2020data} have been shown to support arbitrary dataflows with full-range of TOPS (\secref{sec:dataflow_terminology}), specifically flexible parallelism and shape. However, they suffer from scalability issues due to their all-to-all NoCs.

This work tries to marry the best of both styles. We find that the all-to-all reduction networks in prior works~\cite{kwon2018maeri,sigma} come with prohibitive resource overheads because of redundant reduction paths. This is to accommodate \textit{arbitrary sized reductions}. 
In contrast, \Accel's Neural Engine enables all rows of the 2D PE array to share the same reduction network in a time-multiplexing manner (thereby reducing its cost), without compromising flexibility, throughput, or utilization. 

\label{sec:nest_computation_unit_phase1}

\begin{figure*}[t!]
    \centering
\includegraphics[width=\linewidth]{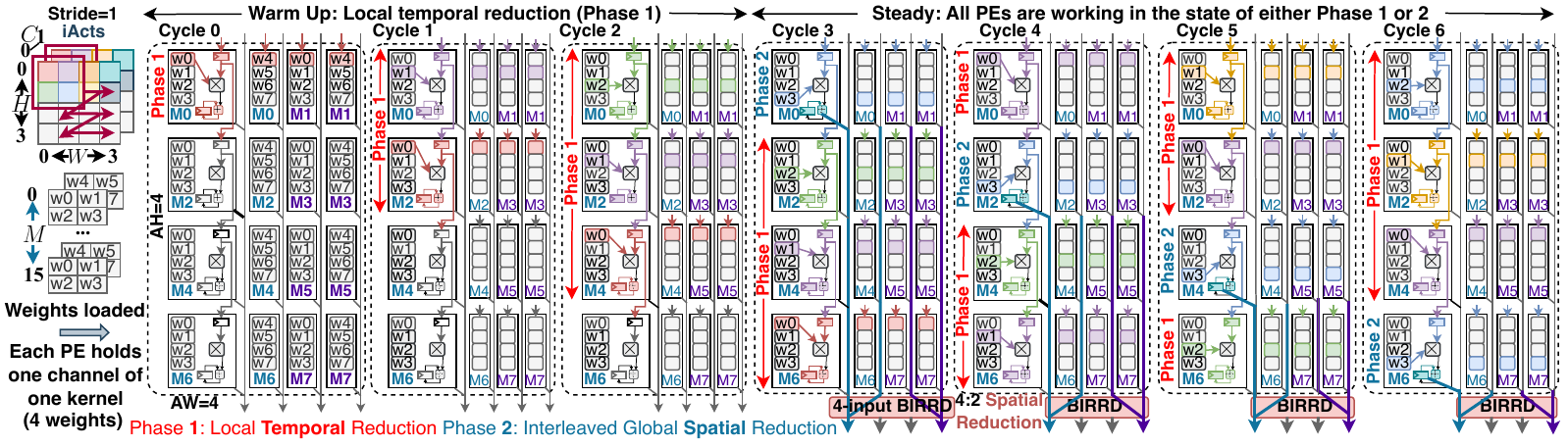}
    \caption{Illustration of the \Accel with \nest and \noc employing a convolutional operation with a $2\times2$ weights featuring 2 input channels ($C=2$) and generating 16 output channels ($M=16$) across a $4\times4$ iAct with 2 input channels.
    The depicted dataflow utilizes a weight-stationary approach, where each PE has a local register file containing a channel of weights ($2\times2$). The dataflow is parallelized for two input channel and two output channel across four PE columns, and for four kernels across four PE rows.
    In each row, four PEs generate 4 partial sums, contributing to 2 final sums, which thus necessitates a 4:2 spatial reduction in the \noc to produce two outputs.
    We assume the weights are already preloaded into \nest before the first cycle in this illustration. 
    The iActs are streamed from the top, undergo multiplication with corresponding weight values (e.g., $w0$ in the top-left PE at cycle-0), and are locally accumulated for the next set of inputs (e.g., until cycle-3 in the top-left PE). 
    Following this initial phase of local temporal reduction, the top row transmits the locally reduced result to the \noc for the second phase of spatial reduction.
    In the steady state, \noc reduces data from one \nest row per cycle (cycles 4-6).
    {\it In steady state, all PEs are working and there is no output bus conflict for PEs of the same column. This is because, during phase-2 of spatial reduction in one PE, remaining PEs of the same column perform local reduction.} In general, $AW\times AH$ \nest takes $AH^2$ cycles to load weights, and ping-pong local registers are instantiated to hide such latency behind computation. \noc could reduce results from PEs at different rows as long as only one PE per column uses the output bus. \textbf{Takeaway:} NEST utilizes local temporal and global spatial reduction to (i) ensure all PEs of the same column share the same output bus without competition while achieving full utilization, and (ii) hide weight loading latency in steady phase.}
    \label{fig:lambda_walk_through_example}
\vspace{-2mm}
\end{figure*}

Specifically, \Accel's \underline{\textbf{N}}eural \underline{\textbf{E}}ngine with \underline{\textbf{S}}patial forwarding and \underline{\textbf{T}}emporal reduction (NEST) works in two phases. One walk-through example for convolution is shown in \figref{fig:lambda_walk_through_example}.

\textbf{Phase 1: Local Temporal Reduction.}
\nest involves local registers in each PE for temporal (local) reduction of partial sums. This is then followed by a phase of global reduction via the reduction network (described in \secref{sec:afft}).

\textbf{Phase 2: Interleaved Spatial Forwarding and Reduction.}
However, unlike prior works where all PEs participate simultaneously in the spatial reduction, the PE rows in \Accel perform spatial reduction one after another, temporally multiplexing on the reduction network. 
Further, while each PE row sends its locally reduced results to the reduction network, PEs in other rows continue computation and reduction locally. 
This is ensured via a pipelining mechanism that guarantees that each row performs $AH$ number of local reductions, before participating in the global reduction.

\textbf{Flexible Dataflow:} \Accel retains 
the ability to support arbitrary dataflow parallelism strategies and shapes (\secref{sec:dataflow_terminology}).
This is because Phase 2 can be configured to create arbitrary-sized reduction groups (i.e., all outputs can be unique or any combinations can be reduced) enhancing mapping flexibility.

\Accel supports inter-layer pipelining. We deploy distinct computation engines for ReLU, BatchNorm, and MaxPooling. For AvgPooling layers, they are transformed into convolution operations and executed within the \nest. When there is a sole requirement for reorder and reduction, the PE Array can be bypassed, directing inputs from \nest directly to the \noc. To optimize storage utilization and reduce data movement costs, all computation engines utilize the same on-chip storage.

\subsection{\Accel's Reordering/Reduction Network -- \noc}
\label{sec:afft}

The Butterfly Interconnect for Reduction and Reordering in Dataflows (\noc) is a multi-stage network designed to reorganize data during the reduction phase. It receives computation results from the previous stage and directs them to new positions in the output buffer while concurrently reducing the data. This process aligns the data in the format needed for the subsequent dataflow, enabling \Accel to seamlessly co-switch (dataflow, layout) for each layer.

\begin{figure*}[t!]
    \centering
    \includegraphics[width=\linewidth]{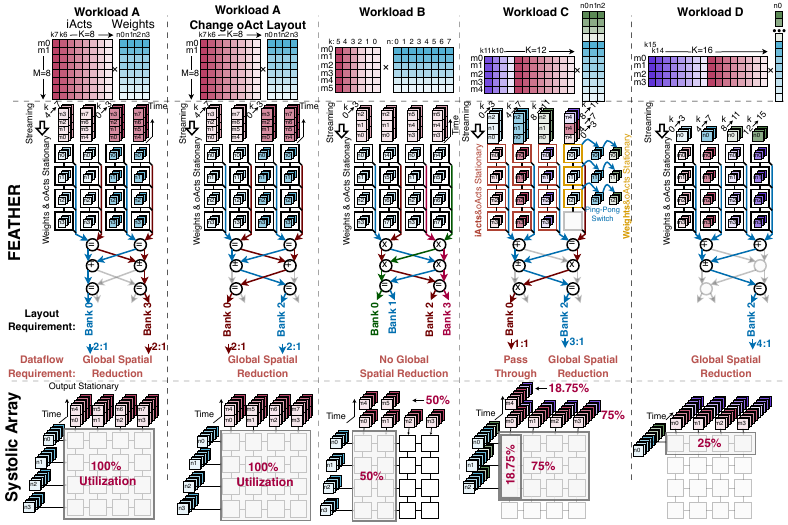}
    \caption{Comparison between per-layer flexible dataflows in \Accel and fixed-dataflow in the systolic array under GEMM. \Accel dynamically alters layout by redirecting oActs to various banks with distinct writing addresses, exemplified by rerouting a blue result from bank 0 (Workload A) to bank 2 (Workload A Change oAct Layout). \Accel consistently outperforms SA in irregular-sized GEMM (Workload B, C, D), achieving near full utilization. Enhanced utilization arises from (1) enabling cross-column spatial reduction using \noc in \Accel, e.g. \Accel maps K dimension across the entire 2D array instead of a single PE in SA under workload D. (2) Eliminating SA's horizontal rigid reuse links, thereby enabling independent mappings across columns, e.g. (Workload C) adopting iAct stationary in first three columns and weights stationary in the last column. \noc could perform pure reordering to change the layout when no spatial reduction is required (e.g. \noc reordering all incoming results to target banks directly under workload B). \textbf{Takeaway:} BIRRD's flexible reduction enhances compute utilization across diverse skewed shapes, expanding the range of dataflows that NEST can efficiently support.}
    \vspace{-2mm}
    \label{fig:lambda_mappings}
\end{figure*}

\subsubsection{\noc Topology} 
\begin{algorithm}
\caption{Inter-stage Connectivity for $AW$-input \noc}
\label{alg:afft_topology} 
\begin{algorithmic}[1]
\setlength{\itemindent}{-3mm}
\setlength{\algorithmicindent}{-3mm}
\STATE \hspace{-2mm} output[$i$][$id$]/input[$i$][$id$] ($id\in[0,AW)$) refers to $id$-th output/input port of BIRRD switches at the stage $i$.
\STATE \textbf{FUNCTION} reverse\_bits(data, bit\_range)
\STATE \hspace{2mm} mask = (1$\ll$bit\_range) - 1
\STATE \hspace{2mm} reversed\_bits = 0
\STATE \hspace{2mm} \textbf{for} $i$ FROM 0 TO bit\_range - 1
\STATE \hspace{4mm} \textbf{if} (data \& (1$\ll$i))
\STATE \hspace{6mm} reversed\_bits $\vert$= (1$\ll$(bit\_range - 1 - i))
\STATE \hspace{2mm} \textbf{return} (data \& $\sim$mask) $\vert$ reversed\_bits\\ 
\STATE \textbf{ENDFUNCTION}
\STATE \hspace{1mm} \textbf{for} $i$ in [0, $2\times log_2(AW)$) // $i$ is stage\_id
\STATE \hspace{2mm} \textbf{for} $j$ in [0, $AW$) // j is port\_id
\STATE \hspace{3mm} output[{\small$i$}][{\small$j$}]$-$input[{\small$i+1$}][reverse\_bits($j$, $\min(log_2(AW), 2+i, 2\times log2(AW)-i)$)]  ($-$ indicates output connects to input)
\end{algorithmic}
\end{algorithm}

The \noc topology is interfaced with \nest engine one side and output buffer on the other side, and is composed of two butterfly networks back-to-back with $log(AW)$-bit bit reverse connections~\cite{dally2004principles}.
This topology grants symmetry with respect to the middle, enabling the construction of each half separately. Each input of \noc receives data from one column-wise bus of the \nest while each output of \noc forwards the result to one output buffer and eventually back to one bank of stationary buffer (StaB, refer to \figref{fig:lambda_arch}) . For \nest with $AW$ columns in total ($AW$ must be a power of 2), the \noc encompasses $2\times log(AW)$ stages\footnote{4-input \noc is a special case with only $2\times log(AW)-1=3$ stages, i.e. the last stages of two half butterfly networks get merged into a single stage.} with $AW/2$ switches located at every stage. The inter-stage connections of \noc are outlined in Alg.~\ref {alg:afft_topology}.

The topology of \noc has been proven to be strictly non-blocking for unicast (any single data point among concurrent inputs sent to a single output)~\cite{arora1990line} and rearrangeably non-blocking for multicasting (at least one data point among all concurrent inputs sent to multiple output ports)~\cite{FatTree,fat_tree_routing,dally2004principles}. We found no multicasting case that it cannot accommodate.

\subsubsection{\noc Reorder-Reduction Switch} 
The \noc is built on 2-input$\times$2-output switch (which we call \textit{Egg}) with adder as shown in~\figref{fig:lambda_micro_arch}.
Each \textit{Egg} is governed by a 2-bit configuration word, allowing for control of four reorder-in-reduction functionalities (shown in~\figref{fig:lambda_micro_arch}) as follows.
\squishlist
\item \textbf{Pass ($=$) / Swap ($\times$)}: directly pass left (right) input data to left (right) output port, or swap them.  
\item \textbf{Add-Left ($\mp$) / Add-Right ($\pm$)}: Accumulates data from input ports and transmits results to the left/right output port, with the secondary output inheriting the input from the same direction.
\squishend
Extra broadcast functions could be added in the Eggs to duplicate accumulated results in multiple banks of StaB.

\subsubsection{\noc Capability and Routing} \noc supports
\squishlist
\item \textit{Arbitrary Reduction}: We define ``reduction group" as a group of inputs that get reduced into one output. $AW$-input \noc supports arbitrary number of reduction groups (up to $AW$).
\item \textit{Arbitrary Reordering}: The rearrangeably multicasting capability enables \noc to route results from many reduction groups to many arbitrary output ports concurrently.
\squishend
The examples of \noc supporting various reordering and reduction patterns are shown in \figref{fig:lambda_mappings}.

From a routing perspective, \textit{reduction can be viewed as a reverse multicasting operation}, where multiple input data points target the same output port and are reduced upon encountering each other at \noc Eggs. 
Thus, we adopt the multicasting routing algorithm~\cite{BENES_Configuration_words} to establish paths and configurations for \noc Eggs, enabling reordering during reduction. If a certain input-output connection cannot be established by the algorithm~\cite{BENES_Configuration_words}, we will brute force all possible configurations.
\figref{fig:lambda_mappings} showcases how \noc supports arbitrary dataflows and layout switching requirements.

\subsubsection{Microarchitectural Benefits of \noc}
\label{sec:rir_benefits}
Generally, distribution networks like Benes in SIGMA~\cite{sigma} or fat-tree in MAERI~\cite{kwon2018maeri} necessitate unicast or multicast capabilities to direct data from relevant on-chip buffer banks to specific processing elements (PEs). This necessity becomes obsolete with \noc (via \rir), as it harmonizes data layouts to coincide with dataflows. 
This enables \Accel to utilize a straightforward point-to-point connection to the input ports of \nest without sacrificing flexibility. 
Consequently, \noc simplifies the requirements for distribution networks in accelerators, thereby minimizing control, resource, and latency expenses.

\subsection{On-chip Storage and Post-processing}
On-chip storage is physically divided into separate buffers with different organizations for concordance with dataflows.

\subsubsection{Stationary (StaB) and Streaming Buffer (StrB)} The typical paradigm of processing convolution or GEMM will keep one type of data stationary, termed a stationary tensor, and stream the other type of data, termed a streaming tensor.
\Accel fetches and processes the streaming tensor in the tile granularity. Both StaB and StrB implement a ping-pong buffer to enable (1) the latency hiding of fetching the next tile from off-chip DRAM, and (2) on-chip inter-layer pipelining.

As for convolution/GEMM (\figref{fig:lambda_micro_arch}), iActs are kept stationary within StaB Ping (or Pong), and the resulting oActs are written back into StaB Pong (or Ping) with a new layout. Meanwhile, weights are streamed via StrB (Ping/Pong).
StaB requires a multi-bank organization ($AW$ banks), with each bank storing a single data piece, to accommodate the varied write addresses in different banks necessitated by layout changes in \Accel. Conversely, StrB adopts a simplified single-bank structure with an $AW$-data bandwidth to conserve area, because weights do not need layout reordering.

\subsubsection{Instruction Buffer (IB)} The configurations for \noc are generated offline and get fetched into IB to configure the reduction networks at run-time.
\subsubsection{Output Buffer (OB)} enables in-situ temporal reduction of partial sums when the reduction size of workloads exceeds the overall reduction capacity of both \nest and \noc. OB has $AW$ banks, and each equipped with a 32-bit adder.
\subsubsection{ZP/Scale Buffer and Quantization Module (QM)} employing quantization schemes from PyTorch FBGEMM~\cite{khudia2021fbgemm} and QNNPACK~\cite{dukhan2018qnnpack}, with 8-bit zero points and 32-bit scales (housed in ZP/Scale Buffer). The quantization module rescaled down 32-bit oActs and then quantized to 8-bit oActs. 

\section{\Accel in Action}
\label{sec:rir}
\vspace{-1mm}

\begin{figure}[t!]
        \centering
        \hspace*{-0.5cm}\includegraphics[width=0.98\columnwidth]{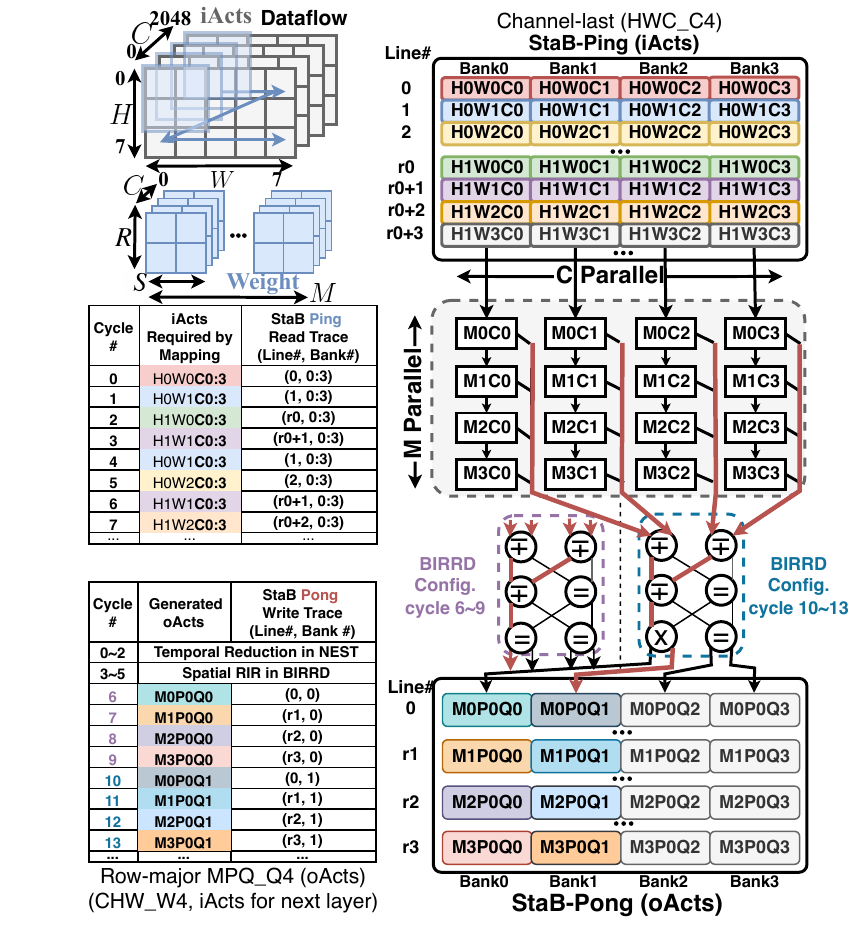}
        \vspace{-3mm}
        \caption{Example of \Accel switching from channel-last layout ($HWC\_C4$) to a row-major format ($MPQ\_Q4(CHW\_W4)$) during reduction without incurring bank conflicts. This is because multiple iActs are reduced into fewer oActs, thereby reducing accesses within each bank. In this example, \nest leverages parallelism along the kernel $M$ and channel $C$ dimensions, reading and vertically streaming four iActs of four input channels from top to bottom. 
        Specifically, at cycle 0, \nest fetches $H0W0C0:3$ from (line $0$, banks $0:3$), as recorded in the StaB Ping read trace. Subsequent cycles involve a two-stage reduction: temporal reduction within the PE for cycles $0$ to $2$, and spatial reduction within \noc for cycles $3$ to $5$, culminating in a single oAct $M0P0Q0$. This oAct is reordered to bank $0$ during reduction and written to line $0$ in the StaB Pong during cycle $6$. 
        \Accel's pipelined processing of following iActs is further exemplified in the read/write trace. $M0:3P0Q0$ target bank $0$ and use connectivity of \noc as shown in the left while \textbf{$M0:3P0Q1$} use the right. For brevity, the notation of $R0:1S0:1$ is omitted, which indicates that each PE in \nest holds four weights of one channel. \textbf{Takeaway:} FEATHER reorders oActs into next layer's desirable layout during reduction, enabling dataflow/layout co-switching.} 
        \label{fig:Full_RIR_Example}
        \vspace{-2mm}
    \end{figure}

In this section, we first showcase one example (\figref{fig:Full_RIR_Example}) of how \Accel leverages \rir to resolve bank conflicts mentioned in \figref{fig:on_chip_reorder} when co-switching dataflow-layout. Then we deep dive into how 
\Accel enables general layout transformations without bank conflicts through two insights.

\subsection{\rir for Bank Conflicts Mitigation and Layout Transform} 
In the example shown in \figref{fig:Full_RIR_Example}, the layout conversion from iActs to oActs is realized via \rir, thereby avoiding the explicit latency in reorder after reduction. 
This efficiency stems from the key insight that \textit{\rir reorders post-reduction oActs into a new layout, rather than directly transforming iActs from one layout to another}. 

Specifically, in the reduction phase, numerous iActs naturally get accumulated into fewer oActs and consequently target fewer banks. For example, four iActs get accumulated to one oAct that targets a single line in \figref{fig:Full_RIR_Example}. Conversely, if we directly transform the layout of iActs from channel-last to row-major, four iActs (H0W0C0:3) would target four different lines within the same bank under row-major layout, leading to bank conflicts.

\subsection{(Dataflow, Layout) Flexibility for Bank Conflicts Eradication} 

While the strategy of `reordering post-reduction oActs' aids in reducing bank conflicts, conflicts may still arise when the number of partial sums to write into memory exceeds the number of writing ports of the memory. This scenario is particularly common in scaled-up $128\times 128$ compute array (Google TPUv4~\cite{tpuv4i}), as it generates more oActs concurrently. 

\Accel fully eliminates conflicts with the second key insight that \textit{\Accel picks the dataflow with the number of oActs (partial sums) matching with the number of memory write ports.} In essence, \Accel employs dataflows free from bank conflicts, and the flexible reduction of the \noc consistently allows \Accel to identify such dataflows with high performance and efficiency. 

In summary, \rir together with flexible dataflows selection enable \Accel to switch among arbitrary layouts without incurring bank conflicts.

\section{Layoutloop}
\label{sec:layoutloop}
\vspace{-1mm}

\Accel enables (dataflow, layout) co-switching at the layer granularity to achieve optimal latency and energy efficiency. However, deciding which (dataflow, layout) to use for \Accel is not trivial because both dataflow and layout have huge space, e.g. $10^{36}\times10^8$ for a single convolution layer (ResNet-50 layer 1)~\cite{gamma}\footnote{A flattening of 4 iActs dimensions $(N=1, C=3, H=224, W=224)$ into two nested loop (\figref{fig:NTLR_Layout_example_simple}) introduces $8!=40320$ order possibilities and $(1, 2, 16, 16)$ factorization possibility. The product leads to $10^8$ layout choices.}, necessitating systematic exploration. For this aim, we enhance Timeloop~\cite{parashar2019timeloop}, a state-of-the-art dataflow search framework with (1) physical storage modeling and (2) systematic layout assessment capabilities, and term it as Layoutloop to distinguish it from native Timeloop.
We employ Layoutloop to explore dataflows under various layouts for \Accel, selecting the dataflow-layout pair that minimizes energy delay product for each layer.

\subsection{Physical Storage Modeling}

Layoutloop models physical storage as ($num\_line \times line\_size$) 2D array with ``$conflict\_depth$" specifying number of lines in each bank with the following reasoning.

\textbf{Bank Organizations:} Current storage uses diverse organizations, including 2D/3D with various groupings. Managing these disparate physical organizations can be complex. However, as storage is usually accessed line by line (or block), we can abstract different organizations into a logical $num\_line \times line\_size$ 2D array. This abstraction allows layout modeling to handle these 2D abstract arrays directly, retaining generality without dealing with specific physical organizations.

\textbf{Bank Port Constraints:} Storage comes with an inherent limitation of the total number of ports in each bank. Concurrent read/write operations exceeding available read/write ports lead to bank conflicts. Thus, $conflict\_depth$ is utilized to denote the total number of lines within a single bank.

\subsection{Bank Conflicts Assessment}
\label{sec:layout_represent}
Layoutloop models slowdown by judging whether bank conflicts occur when analyzing data access to the on-chip buffer with a specific layout. A $\max(N_P/N_L, 1)$ slowdown is introduced if $N_L$ lines are accessed from a bank with $N_P$ ports.
Finally, we also modify Timeloop's mapper to consider data layout during dataflow search.

\label{sec:layout}

\section{Evaluation}
\label{sec:evaluation}
\vspace{-1mm}

\begin{table*}[!htp]\centering
\caption{Evaluation setup for SoTAs and \Accel ($\rightarrow$ indicates the modifications from the original design).}
\vspace{-2mm}
\label{tab:evaluation_setup}
\scriptsize
\begin{tabular}{c>{\raggedright\arraybackslash}p{2.5em}>{\centering\arraybackslash}p{3em}>{\raggedleft\arraybackslash}p{9em}>{\raggedright\arraybackslash}p{6em}>{\centering\arraybackslash}p{4.9em}>{\raggedleft\arraybackslash}p{4em}cccc}\hline
& \multicolumn{3}{c}{{Run-time Flexibility in}}  & &  on-chip BW  &   &  \multirow{2}{12em}{Area ($\mu m^2$)/Clock Frequency}  & Evaluation Method \\
& (Layout, & Dataflow\circled{2}, & Reorder) & (\#PE,& bit/cycle, & DataType) &   & Real-device/Layoutloop \\\hline
Edge TPU & (fix ,& T,& none) & (1024,& unknown,& int8) & 500 MHz (ASIC) & Coral USB accelerator~\cite{coral_usb_accel} \\
Xilinx DPU & (fix ,& T,& none) & (1152,&  unknown,&  int8) &  100 MHz (FPGA) &end-to-end on ZCU104 \\
Gemmini & (fix,& T,& none) & (1024,&  512,&  int8) &  50 MHz\circled{1} (FPGA) &FireSim on AWS EC2 F1 \\
\textbf{\Accel} & (flexible,& TOPS,& Reorder In Reduction) & (1296,&  720,&  int8)  & 100 MHz (FPGA) &end-to-end on ZCU104 \\
\hline
NVDLA-like & (fix,& T, & none)  &(16$\times$16,&  25$\rightarrow$256,&  int8) &    808K (TSMC-28, 1 GHz) \circled{5} &Layoutloop \\
Eyeriss-like & (fix,& TS,& none)  &(14$\times$12$\rightarrow$16$\times$16,&  192$\rightarrow$256,&  int16$\rightarrow$int8)  & 1394K (TSMC-65, 200 MHz)  &Layoutloop \\\hdashline
SIGMA-like & (fix\circled{3}, &TOPS, & none)  &  & &  &   & \\
SIGMA-like & (flexible, & TOPS,& off-chip reordering\circled{4})  &  (65536$\rightarrow$256,&  256,&  bf16$\rightarrow$int8) &  990K (TSMC-28, 500 MHz) &Layoutloop \\
Medusa-like & (flexible, & TOPS,& on-chip line rotation)  &  &  &  &  & \\ \hdashline
TPU-like & (flexible, & TO,& transpose/row reorder)  & (8$\times$``128$\times$128", & 256,& int8) & 600K (7nm, 1050 MHz)  &Layoutloop \\ 
MTIA-like & (flexible,  & TOP,& transpose)  &  (64$\times$``32$\times$32",&  1024,& int8) & 373K (TSMC-7, 800 MHz)  & Layoutloop \\ 
\textbf{\Accel} & (flexible,& TOPS,& \rir)  & (16$\times$16,&  8000,&  int8) & 338K (TSMC-28, 500 MHz)  &Layoutloop \\ \hline
\end{tabular}

{\scriptsize \circled{1}: Latency scaled to 100 MHz. We standardized the frequency at 100 MHz just for a fair comparison purpose, which is not indicative of the maximum clock frequency achievable on ASIC implementations.; \circled{2}: Terminology defined in \secref{sec:dataflow_terminology}; \circled{3}: HWC\_C4W8 or HWC\_C32; \circled{4}: off-chip bandwidth=128 GB/s. \circled{5}: compute area only.}
\end{table*}

\subsection{Methodology}
\label{sec:vi_a}
We implement \Accel in Verilog and Xilinx HLS. Verilog-based implementation delivers precise micro-architecture design while HLS-based implementation enables the native usage of Xilinx IPs for buffer, control, and peripherals for better end-to-end performance on Xilinx FPGAs. We evaluate its resources on TSMC 28 nm high performance technology node using the Verilog-based implementation. We compare its end-to-end wall-clock latency against SoTAs with open-sourced end-to-end implementations on real FPGA devices. We also model \Accel in Layoutloop (\secref{sec:layoutloop}), including energy overheads, to compare it against SoTA accelerators that do not have open-sourced end-to-end deployable codes. \tabref{tab:evaluation_setup} summarizes our evaluation setup.

\subsubsection{Baselines and Workloads}\hfill

\textbf{Baselines for real-device} evaluations.
We compare \Accel against Xilinx DPU~\cite{xilinxdpu} , Gemmini~\cite{gemmini}, and Edge TPU~\cite{seshadri2022evaluation}, as they can be deployed in an end-to-end fashion. 
\Accel and Xilinx DPU~\cite{xilinxdpu} are deployed on the same Xilinx ZCU 104 FPGA board. While Gemmini is deployed on AWS-F1 FPGA server~\cite{gemmini} using FireSim to emulate its per-layer processing latency. Edge TPU~\cite{seshadri2022evaluation} runs on a USB accelerator~\cite{coral_usb_accel} attached to a Raspberry Pi 4B. As for all four designs, we normalize throughput by the number of PEs (i.e., MAC units) and clock frequency~\footnote{Both GEMMINI and FEATHER could run at 1 GHz under TSMC 28 nm ASIC flow. However, the parallel simulation synthesis toolchain of firesim limits GEMMINI's clock frequency to 50 MHz on AWS's f1.2xlarge FPGA.} for a fair comparison. 

\textbf{Baselines for Layoutloop.}
\Accel is further compared against NVDLA~\cite{nvdla}, Eyeriss~\cite{eyeriss} and SIGMA~\cite{sigma} in Layoutloop. Detailed modifications/specs are listed in \tabref{tab:evaluation_setup}

\textbf{Workload.} BERT (representative of cloud workloads);  ResNet-50 and MobiletNet-V3 (Mob-V3) as edge workloads.

\subsubsection{\Accel Dataflow/Layout Setup}
\squishlist
\item \textbf{Search Space.} Dataflow design space is constructed by arbitrary nested loops as shown in \figref{fig:workload_dataflow}. We use layout patterns used by prior accelerators~\cite{AI_Accel_survey} as layout space \footnote{Conv: HWC\_C32, HWC\_W32, HWC\_H32, HWC\_C4W8, HWC\_C4H8, HWC\_W4H8, HWC\_C4W4H2; GEMM: we note input/weights/output as $M\times K$/$N\times K$/$M\times N$ with inputs layout as MK\_K32, MK\_M32, MK\_M4K8.}.
\item \textbf{Searching Algorithm.} We exhaustively search layout space for global optimal.
To find optimized dataflows, we use Timeloop’s internal hybrid search algorithm (exhaustive + search-space pruning). Recent works~\cite{kao2022demystifying} show that its results are comparable to sophisticated search methods~\cite{gamma, cosa, MindMappings} but is slower in wall clock time. We ran the search with multiple threads constrained on search size and victory conditions.
\item \textbf{Performance Metric.} We use Energy-Delay-Product (EDP) as the performance metric for a dataflow/layout pair.
\item \textbf{Overall Search Flow.} Dataflow/layout cosearch is conducted for each layer independently. The optimal dataflow-layout pair with the best EDP is chosen for each layer of ResNet-50 and Mob-V3 in the Layoutloop evaluation. For end-to-end FPGA deployment of ResNet-50, we simplify engineering efforts by selecting the two layouts with the best latency and energy efficiency on DepthWise Conv. and typical Conv., and enable \Accel to switch between them per layer.
\squishend

\label{sec:layout_choice}

\begin{figure}[t!]
    \vspace{-2mm}
    \includegraphics[width=\linewidth]{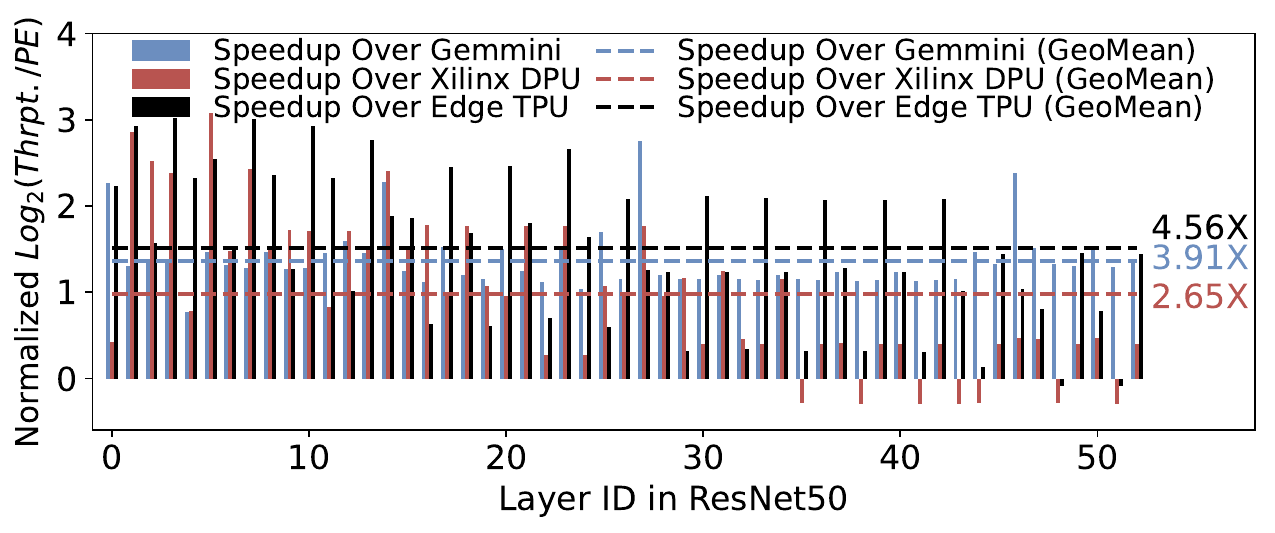}
    \vspace{-2mm}
    \caption{FEATHER vs. SoTAs on real devices. We run each layer for 100 times to obtain average layer latency, and then normalize throughput by number of PE and clock frequency.}
    \vspace{-3mm}
    \label{fig:Latency_Comparison_LAMBDA_DPU}
\end{figure}

\begin{table*}
\begin{minipage}{0.25\linewidth}
    \label{table:student}
    \centering
    \scriptsize
    \begin{tabular}[h]{>{\centering\arraybackslash}p{5.4em}cc}
    \hline
           & \multicolumn{2}{c}{Layout Setup} \\ 
      Label &  compile & runtime \\ \hline
      NVDLA-like & fixed & fixed\\
      Eyeriss-like & fixed & fixed \\
      SIGMA-like & flex & fixed \\
      Medusa-like & \multicolumn{2}{c}{line rotation (\figref{fig:line_rotation})}  \\
      MTIA-like & \multicolumn{2}{c}{transpose (\figref{fig:transpose})} \\
      TPU-like & \multicolumn{2}{c}{\makecell{transpose + row\\ reorder (\figref{fig:row_reorder})}} \\ 
      FEATHER & \multicolumn{2}{c}{arbitrary reorder (\figref{fig:arbitrary_reorder})} \\  \hline 
    \end{tabular}
\end{minipage}\hfill
\begin{minipage}{0.71\linewidth}
    \centering
\vspace{-2mm}
    \includegraphics[width=\linewidth]{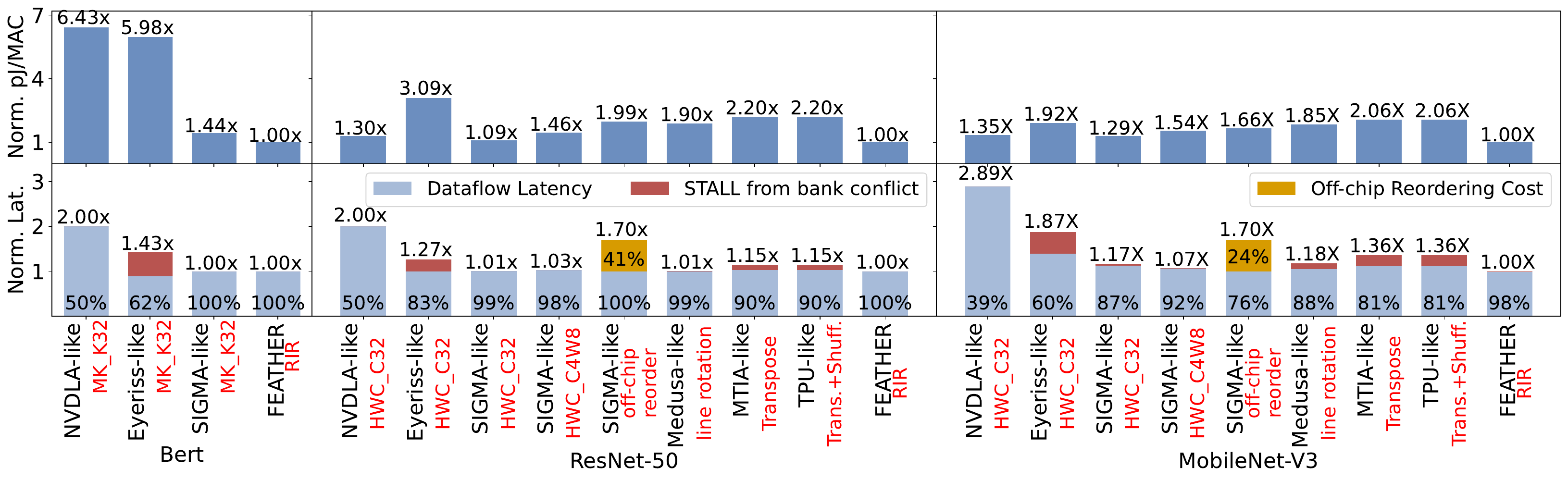}
\end{minipage}
  \vspace{-2mm}
  \captionof{figure}{\Accel vs. SoTA using Layoutloop (Percentage inside each blue bar indicates average steady-state PE utilization. Red bar indicates bank conflict slowdown, while yellow bar indicates off-chip reordering costs. Lower is better. (The red text in the x-axis of the right chart mentions the fixed layout or the layout reordering mechanism for each design.)
  }
  \vspace{-3mm}
   \label{fig:Compare_to_SoTA}
\end{table*}

\subsection{End-to-end Real-device Latency Evaluation}
\subsubsection{\Accel vs. Gemmini}
\Accel achieves a $3.91$$\times$ geomean normalized throughput improvement than Gemmini as shown in \figref{fig:Latency_Comparison_LAMBDA_DPU}, as Gemmini adopts a fixed dataflow (weights stationary with degree of parallelism being 16 
in both C and M), leading to under-utilization when C of workload is not divisible by 16. The flexibility of \Accel in the parallelism of M,C,H,W delivers its performance improvement.

\subsubsection{\Accel vs. Xilinx DPU}
The $2.65$$\times$ more throughput of \Accel over Xilinx DPU stems from the low steady-state utilization of Xilinx DPU under convolution $3\times3$ (75\% utilization), $7\times7$ (21.8$\sim$87.5\% utilization), and \Accel pushes both utilization to 100\% and 90.4\% in the steady state. 
This is because Xilinx's DPU with 1152 PEs only supports a single dataflow with parallelism $(12, 12, 8)$ in $(M, C, H/W)$. 
In deep layers with a large number of input channels (C) and kernels (M), both \Accel and Xilinx DPU achieve a steady utilization of 100\%. 
However, Xilinx DPU outperforms \Accel for these layers as our controller is not as optimized as DPU's (an engineering optimization part of our future work).

\subsubsection{\Accel vs. Edge TPU} \label{sec:edge_tpu_cmp} $4.91\times$ speedup comes from flexibility of FEATHER in dataflow and layout.

\subsection{Layoutloop-based Latency/Energy Evaluation}
\label{sec:layout_loop_res}
Latency and energy efficiency are affected by (1) compute utilization determined by dataflows, and (2) effective memory bandwidth considering bank conflicts. 

With per-layer dataflow-layout switching, \Accel achieves the peak steady-state utilization of 100\%, 100\%, and 98.3\% with zero bank conflict slowdown under BERT, ResNet-50, and Mob-V3, separately. This indicates that FEATHER consistently provides desirable layout for all three workloads.

\subsubsection{\Accel vs. NVDLA}
\Accel achieves 2$\times$/2$\times$/2.89$\times$ speedup and 6.43$\times$/1.3$\times$/1.35$\times$ higher efficiency over NVDLA under BERT/ResNet-50/Mob-V3. Leveraging fixed weights/output stationary dataflows under a fixed HWC\_C32 layout, NVDLA will not encounter any bank conflicts, which delivers good energy efficiency. However, NVDLA only allows flexible tiling sizes $T$ (definition in \secref{sec:dataflow_terminology}), and such dataflows suffer from low utilization (50\% and 39\%), explaining higher normalized latency in \figref{fig:Compare_to_SoTA}.

\subsubsection{\Accel vs. Eyeriss} The 1.43$\times$/1.27$\times$/1.87$\times$ speedup and 5.98$\times$/3.09$\times$/1.92$\times$ higher efficiency of \Accel over Eyeriss comes from both bank conflicts elimination and higher-performance dataflows. Specifically, Eyeriss adopts row-stationary dataflow and enables flexible tiling and shape but such sub-flexibility comes with the price of bank conflicts. Compared against Eyeriss-like, \Accel incurs $6$\% more area by introducing \noc and controller, as shown in \figref{fig:over_resource_breakdown}. 

\subsubsection{\Accel vs. SIGMA} 
\label{rev:prior_fgpa_accel}

SIGMA supports flexibility in all 4 dimensions of dataflows (\secref{sec:dataflow_terminology}), but does not have reordering capability (\secref{sec:reorder_analysis}). We thus evaluate static layout, off-chip and three on-chip reordering scenarios, as illustrated in \figref{fig:reorder_overview}.
\squishlist
\item \textbf{Fixed Layout + No Reordering:} SIGMA adopts a fixed layout at runtime and keeps iActs and oActs always on-chip. And we leverage Layoutloop to search dataflows with minimal bank conflicts. Results under two layouts (HWC\_C32 and HWC\_C4W8) out of seven layouts delivering relatively better latency and energy efficiency are depicted in \figref{fig:Compare_to_SoTA}.
Proposed Layoutloop could fully utilize SIGMA's flexibility in TOPS to identify dataflows with fewer bank conflicts, resulting in a small speedup of \Accel over SIGMA (the same for BERT, 1.01$\times$/1.03$\times$ for ResNet-50, and 1.17$\times$/1.07$\times$ for Mob-V3).

Although Layoutloop could identify dataflows for SIGMA with good latency, such dataflows always need to read more lines than dataflows adopted by \Accel, resulting in the high energy efficiency of \Accel (1.44$\times$ for BERT, 1.09$\times$/1.46$\times$ for ResNet-50, and 1.29$\times$/1.54$\times$ for Mob-V3).

\item \textbf{Concordant Layout + Off-chip Reordering:} SIGMA pays the latency and energy costs to send oActs back off-chip and changes layout per layer. Such reordering cost is explicitly shown in \figref{fig:Compare_to_SoTA}. In the case of high compute-intensive ResNet-50, the latency of off-chip reordering could be almost hidden behind computation latency when adopting HBM with 128 GB/s, leading to the pure energy costs of moving data back and forth between HBM and compute. By contrast, in low compute-intensive MobV3, off-chip reordering exposes 24\% critical latency, which further restricts the performance of dataflows as SIGMA has to use some dataflows with the least off-chip accesses. This explains the 1.7$\times$/1.7$\times$ speedup and 1.99$\times$/1.66$\times$ efficiency improvement of \Accel.
\item \textbf{Flexible Layout + On-chip line Rotation:} SIGMA is equipped with line rotation, proposed in Medusa~\cite{Medusa}, to mitigate bank conflicts when reading three lines from the same bank. But typical workloads often access more than $4$ lines per cycle. Further, all seven on-chip layouts utilized in the paper require word-granularity data reordering to switch from one to the other, a capability supported by \Accel but not line rotation, which explains the 1.01$\times$/1.18$\times$ speedup and 1.90$\times$/1.85$\times$ efficiency improvement of \Accel. 
\item \textbf{Flexible Layout + On-chip Transpose (MTIA-like):} We enhance SIGMA with on-chip data transpose (\figref{fig:transpose}), the reordering capability provided by MTIA and TPUv4. Transpose is effective for reducing bank conflicts caused by single-dimensional parallelism. However, multi-dimensional parallelism based bank conflicts require finer-grain data reordering, a function supported by \rir but not by transpose. Therefore, FEATHER demonstrates speedups of 1.15$\times$/1.36$\times$ and achieves 2.2$\times$/2.06$\times$ greater efficiency, highlighting its superior handling of complex data layout transformations.
\item \textbf{Flexible Layout + On-chip Transpose and Row-reorder (TPU-like):} On top of MTIA-like, we further add row-reorder (\figref{fig:row_reorder}). Yet, this enhancement does not reduce on-chip buffer accesses nor modify data locations compared to transpose alone, resulting in no further latency saving or efficiency gains. 
\squishend

\vspace{-2mm}
\subsection{Resources and Timing Evaluation under TSMC 28nm}
\vspace{-1mm}

\subsubsection{\noc vs. FAN~\cite{kwon2018maeri}/ART~\cite{sigma}}

We have implemented \noc in Verilog, and obtained its post-layout resources under different scales.
\figref{fig:reduction_noc_evaluation} provides a comparative evaluation of \noc with other reduction networks like SIGMA's FAN and MAERI's ART, considering int32 adders. Here, the area is represented by lines while the bars demonstrate power consumption. An $AW$-input \noc has more stages ($2log_2(AW)$) than FAN/ART ($log_2(AW)-1$). Consequently, \noc consumes about 1.43$\times$/2.21$\times$ more area and 1.17$\times$/2.07$\times$ more power than FAN/ART. Despite these overheads, the adoption of \noc is justifiable for two primary reasons:
\squishlist
\item  Unlike FAN/ART which requires one $AW\times AH$-input instance for all 1D PEs, a single $AW$-input \noc instance can fulfill the reordering and reduction needs for all 2D PEs. As a result, when integrated into \Accel, \noc achieves a resource-saving of $94\%$ over the FAN in SIGMA, as indicated by the Reduction NoC (Redn. NoC in \figref{fig:over_resource_breakdown}).
\item While both MAERI and SIGMA necessitate a complex distribution NoC such as fat-tree, crossbar, or Benes, \noc eliminates such requirements as data always come in a perfect layout without further redistribution demands (\secref{sec:rir_benefits}). Thus \Accel replaces  distribution NoC with pt-to-pt connections.
\squishend

\subsubsection{\Accel vs. SIGMA/NVDLA}
The combined simplification of the distribution NoC and implementation of a singular \noc instance results in a substantial $2.93$$\times$ resource reduction of SIGMA - for \Accel with an equal number of 256 PEs, as illustrated in \figref{fig:over_resource_breakdown}. \Accel has large local memory as each PE in $AW\times AH$ \Accel needs to keep sufficient data inside local memory to perform local reduction when other PE rows are using oAct buses. Further, \Accel adopts 2D PE array with better scalability compared with 1D PE design in SIGMA. 
We further implement a NVDLA-like 1D PE array serving as a fix-dataflow baseline. 

\subsection{Timing Analysis}
\label{sec:timing_analysis}

\begin{figure}[t]
    \centering
    \vspace{-2mm}
     \subfloat[Reduction Network. \label{fig:reduction_noc_evaluation}]{{\includegraphics[width=0.515\columnwidth]{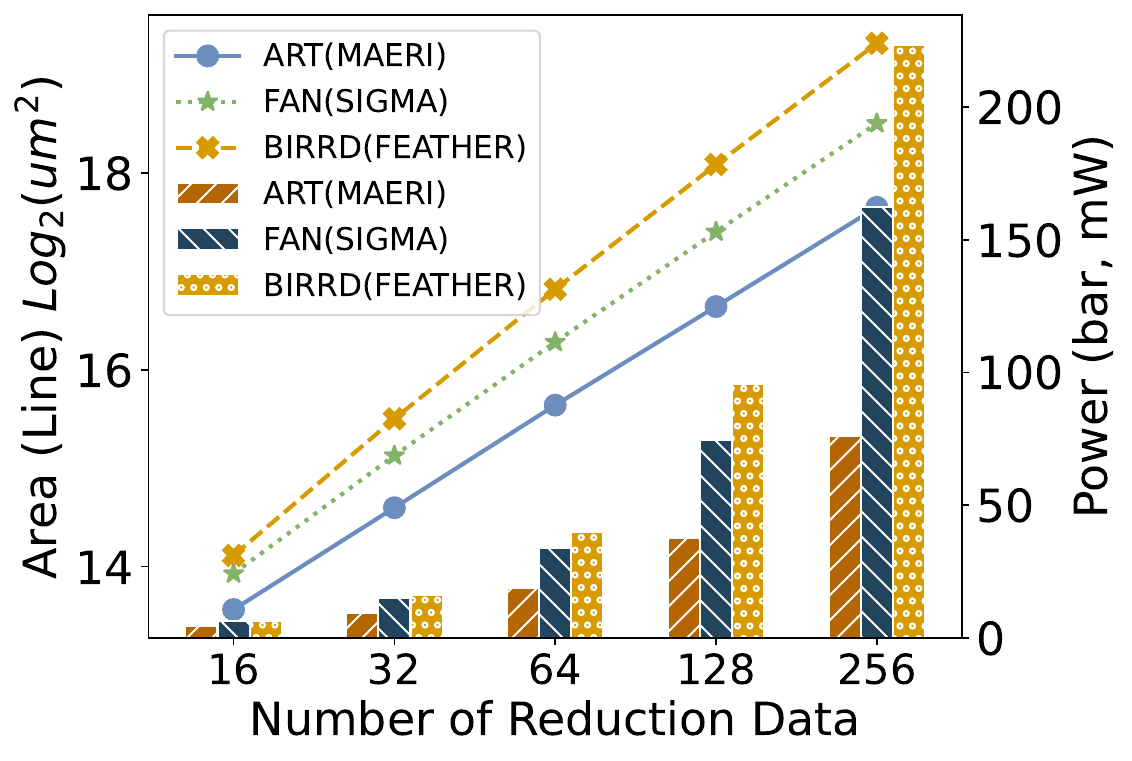}}}
    \subfloat[Resources Breakdown. \label{fig:over_resource_breakdown}]{{\includegraphics[width=0.485\columnwidth]{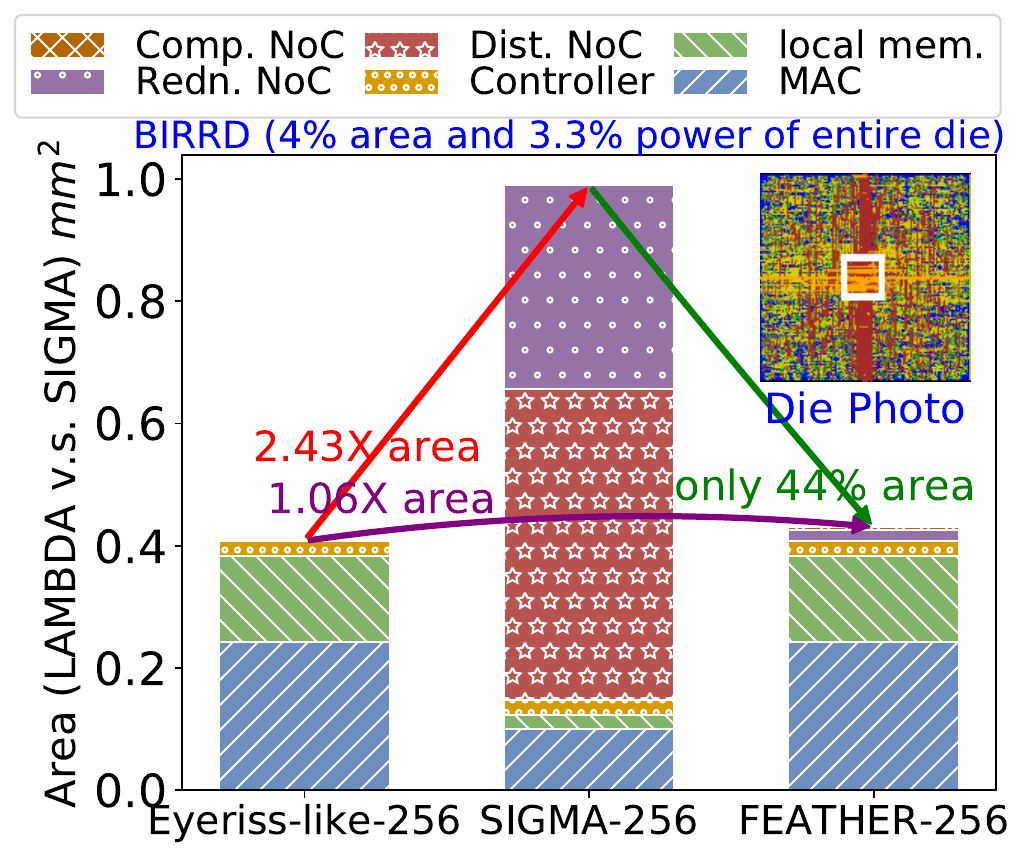}}} 
    \vspace{-2mm}
\caption{ASIC resource comparison (\Accel vs. SoTA). 16$\times$16 \Accel place-and-route at TSMC 28nm.}
    \vspace{-3mm}
    \label{fig:resource_evaluation}
\end{figure}

We layout \Accel with 64, 256, and 1024 PEs, requiring \noc with 8, 16, and 32 inputs. The die photo of \Accel with $16\times 16$ PEs is shown in \figref{fig:over_resource_breakdown} revealing that \noc consumes only 4\% of the overall post-layout area in the TSMC 28nm process. \noc does not have long wires because it is placed outside the PE array as a standalone module instead of spreading among PEs like FAN in SIGMA or ART in MAERI. \noc attains a peak clock frequency of 1.5 GHz across all scales. The timing critical path of \Accel is the wire connecting local weights registers to 9-bit multiplier in PE, maxing at 1 GHz, similar to SoTA accelerators~\cite{tpuv4i}.

\section{Related Work}
\label{sec:relatedWork}
\vspace{-1mm}

\textbf{DNN Accelerators.} Most DNN accelerators today (especially those with end-to-end deployment support~\cite{xilinx_dpu,gemmini,nvdla,tpuv4i,xilinx_AIE,behnam2023subgraph,micro_SUSHI})
either rely on fixed dataflows - thus fixed layouts or support flexible dataflows~\cite{chen2018eyeriss,kwon2018maeri,sigma,DSAGen} but do not consider effects of data layout when switching dataflows, both of which hurt performance. \tabref{tab:cmp_other_accel},\ref{tab:comparision_reordering} contrast these.

\textbf{Layout Reordering Support.} To mitigate bank conflicts, Medusa~\cite{Medusa} introduces line rotation, which rotates one line inside the conflicted bank into different banks when moving data from off-chip DRAM to on-chip compute. However, typical accelerators with higher parallelism require data in finer word granularity, which leads bank conflicts to line rotation. 

To enable word-level layout reordering, MTIA~\cite{MITA} introduces a Memory Layout Unit (MLU) that enables transposing, concatenating, and reshaping with 4/8/16/32-bit data types. Besides these three layout transforms, \Accel further supports arbitrary data reordering layout transformations through \noc. Moreover, extra reordering latency from MLU is completely hidden behind reduction in \Accel through RIR.

\section{Conclusion}
\vspace{-1mm}

This work motivates the need for data layout reordering support for switching between dataflows in DNN accelerators. We introduce \Accel, an innovative accelerator incorporating a novel multi-stage reduction networks called \noc to implement a unique on-chip reordering strategy for reordering data in reduction. This facilitates simultaneous dataflow and layout switching in \Accel without explicit latency costs, resulting in speedup over SoTAs by using less area.

\section{Acknowledgement}
\label{sec:acknowledge}

We thank Hyoukjun Kwon, Geonhwa Jeong and Raveesh Garg for feedbacks, Angshuman Parashar for concordant layout terminology, Sixu Li for place-and-routing, Yongan Zhang for ZCU104 maintenance. This work was supported in part by ACE, one of seven centers in JUMP 2.0, a Semiconductor Research Corporation (SRC) program sponsored by DARPA.
\appendix  
\subsection{Abstract} This artifact contains three different flows to evaluate FEATHER with different fidelity against different baselines:  (1) end-to-end evaluation on the realistic ZCU 104 FPGA evaluation board, (2) analytic analysis using the mapping search and evaluation framework, LayoutLoop (details in \secref{sec:layoutloop}) and (3) Verilog implementation with synthesis and place-and-routing evaluation flow.

\subsection{Artifact check-list (meta-information)}

Inside the repo, we provide three different experiments in three separate folders. Each folder consists of (1) pre-run results for each experiments, and (2) detailed step-by-step operation to reproduce the pre-run results.

\subsubsection{Experiment Set 1 - \figref{fig:Latency_Comparison_LAMBDA_DPU}}
deploys \Accel on ZCU 104 FPGA board and evaluate the end-to-end layer-wise latency of processing ResNet-50.
\squishlist
\item Pre-run per-layer results for inference convolution 3$\times$3, 1$\times$1 layers of ResNet50 are stored in the output of provided jupyter notebook ``feather.ipynb". 
\item Pre-built FPGA bitstream for running on ZCU104 FPGA.
\squishend

\subsubsection{Experiment Set 2 - \figref{fig:Compare_to_SoTA}} automated dataflow-layout co-exploration, leveraging Layoutloop, to investigate performance of various baselines and \Accel.
\squishlist
\item LayoutLoop Framework in path ``LayoutLoop/layoutloop".
\item Configurations for FEATHER (arbitrary layout choice), SIGMA (arbitrary layout choice), SIGMA (off-chip reordering), MTIA-like (Transpose), TPU-like (Transpose + Shift), SIGMA-like (HWC\_C4W8), SIGMA-like (HWC\_C32), Medusa-like (Line Rotation), Eyeriss-like (HWC\_C32), NVDLA-like (HWC\_C32), at path ``LayoutLoop/configurations".
\item Pre-run results at path ``LayoutLoop/pre\_run\_results".
\item Dataflow-Layout Co-searching scripts ($>$ 24 hours runtime).
\squishend

\subsubsection{Experiment Set 3 - \figref{fig:resource_evaluation}}
ASIC synthesis and place-and-route on Verilog implementation of \Accel.
\squishlist
\item Pre-run results.
\item automated scripts for synthesis and PnR ($>$ 96 hours).
\squishend

An overview of the dependency is listed as follows.
{\small
\begin{itemize}
  \item {\bf Algorithm}: We adopt pruned random searching algorithm to explore large dataflow design space under various layouts.
  \item {\bf Program}: We provide three experiment sets with click-and-run program entries.
  \item {\bf Compilation}: We provide step-by-step compilation guideline in the repo and a pre-built docker with all pre-compiled programs.
  \item {\bf Binary}: The pre-built hardware binary for realistic FPGA deployment is provided in the repo.
  \item {\bf Model:} We use ResNet-50 for end-to-end FPGA evaluation, and meta data from ResNet-50, MobileNet-v3 and BERT for LayoutLoop analytic analysis.
  \item {\bf Run-time environment: } Ubuntu, version does not matter.
  \item {\bf Hardware: } We provide the access for ZCU 104 evaluation board. Users need to have multi-core CPU with 32 GB memory.
  \item {\bf Metrics: } Latency, Latency-Energy-Product
  \item {\bf How much time is needed to prepare workflow (approximately)?: } Estimated (1) 1 minutes (2) 10 mins (3) 10 mins.
  \item {\bf How much time is needed to complete experiments (approximately)?: } Estimated (1) 10 minutes (2 \& 3) $>$ 96 hours.
  \item {\bf Publicly available?: } Yes
  \item {\bf Code licenses (if publicly available)?: } MIT
  \item {\bf Workflow framework used?: } Our proposed Layoutloop is built upon Timeloop~\cite{parashar2019timeloop}
\end{itemize}
}

\subsection{Description}

\subsubsection{How to access}
{\em The artifact is available at \url{10.5281/zenodo.10999154}.}

\subsubsection{Hardware dependencies}
ZCU104 FPGA evaluation board is required to reproduce the end-to-end evaluation results of FEATHER under ResNet-50.

\subsubsection{Software dependencies}
\squishlist
\item (Experiment-1) Prebuilt PYNQ 3.0.1 image for ZCU104 FPGA board with Python 3.10.2, which is available at \url{http://www.pynq.io/boards.html}.
\item (Experiment-2) scons, libconfig++-dev, libboost-dev, libboost-iostreams-dev, libboost-serialization-dev, libyaml-cpp-dev, libncurses-dev, libtinfo-dev, libgpm-dev, cmake; Python 3.8 with matplotlib, numpy, pandas. 
\item (Experiment-3) Synopsys 2022.12-SP5 and Cadence innovus v21.14-s109\_1, both could be other versions. TSMC 28nm technology standard cell libraries.
\squishend

\subsection{Installation}
We provide detailed installation guide and step-by-step instructions to reproduce the results in the repository. 
\subsubsection{Results Visualization}
Our visualization requires the following python packages.
\begin{python}
$ git clone <repo_url>
$ conda create -n <favoriate_name> python=3.8
$ conda activate <favoriate_name>
$ pip3 install matplotlib numpy pandas
$ python results_generation.py
\end{python}

\subsubsection{LayoutLoop Setup} The compilation and code base requires following libraries on Ubuntu system.
\begin{python}
$ sudo apt install scons libconfig++-dev \
libboost-dev libboost-iostreams-dev \
libtinfo-dev libboost-serialization-dev cmake \
libyaml-cpp-dev libncurses-dev libgpm-dev 
\end{python}

We also provide pre-built docker at \url{https://tinyurl.com/layoutloop}. Install the following packages to run the docker.
\begin{python}
$ sudo apt-get install docker-ce containerd.io \ 
docker-ce-cli  docker-buildx-plugin \ 
docker-compose-plugin
$docker load -i feather_layoutloop_docker.tar.gz 
\end{python}

\subsection{Experiment workflow}

\subsubsection{Experiment Set 1, End-to-end Inference on FPGA}
\squishlist
\item We require ZCU104 with pre-built PYNQ image, which provides a jupyter notebook portal.
\item Copy the provided ``feather/feather.ipynb" into the jupyter notebook and then run all blocks.
\squishend

\subsubsection{Experiment Set 2, LayoutLoop Analytic Analysis} run the following commands within the pre-built docker. The DSE might take 1 day to finish, depending on the machine.

\begin{python}
$ docker run -it feather_layoutloop
$ git clone <provided_url>
$ cd FEATHER/LayoutLoop/configurations
$ git pull
$ make clean
$ make conv_dse # DSE for ResNet-50, Mob-v3
$ make gemm_dse # DSE for Bert
\end{python}

\subsubsection{Experiment Set 3, Synthesis and PNR under TSMC 28nm} For synthesis, we use design compiler ``dc\_shell"

\begin{python}
<setup environment for synopsys>
$ cd FEATHER/FEATHER_RTL/scripts/
$ source :run_syn
\end{python}

For place and routing, we use innovus.

\begin{python}
<setup environment for innovus>
<Finish Synthesis First>
$ cd FEATHER/FEATHER_RTL/
$ innovus
> source PnR.tcl
\end{python}

\subsection{Evaluation and expected results}

\subsubsection{Exp. Set 1: End-to-end latency on FPGA} The layerwise latency of running various models will be shown in the end of the jupyter notebook. We normalize results from different designs using ``normalized throughput per PE", where throughput is measured by inverse of latency under single batch. The visualized result is shown in \figref{fig:Latency_Comparison_LAMBDA_DPU}.

\subsubsection{Exp. Set 2: LayoutLoop Analytic Analysis} Per-layer results from Layoutloop could be found at ``LayoutLoop/configurations/results" with following naming pattern.

\squishlist 
\item {design\_name}\_{layout\_policy}\_slowdown.csv
\item {design\_name}\_{layout\_policy}\_utilization.csv
\item {design\_name}\_{layout\_policy}\_pj\_commpute.csv
\item {design\_name}\_{layout\_policy}\_cycle.csv
\squishend

We calculate GeoMean of ``pJ/compute" and ``cycle", and then normalize all results by FEATHER's performance with the visualized results shown as \figref{fig:Compare_to_SoTA}.

\subsubsection{Exp. Set 3: Synthesis and PNR under TSMC 28nm}
The final reports of synthesizing \Accel at a specific scale will be listed in the ``reports" folder, including 
\squishlist 
\item feather\_top\_area.rpt
\item feather\_top\_dw\_area.rpt
\item feather\_top\_power.rpt
\item feather\_top\_timing.rpt
\squishend

The final reports of PnR contain 
\squishlist 
\item area.rpt, which contains Post-PnR area value.
\item power.rpt, which contains Post-PnR power value.
\item time, timingReports. \# Both are timing reports.
\squishend

\begin{table}[!ht]
    \centering
    \caption{Post-PnR \Accel Area/Power at various shapes.}
    \label{tab:resources}
    \begin{tabular}{cccc}
    \hline
        Shape & Area ($\rm \mu m^2$) & Power($\rm mW$) & Frequency (GHz) \\ \hline
        64$\times$128 & 36920519.69 & 26400.00 & 1.00 \\ 
        64$\times$64 & 18389176.19 & 13200.00 & 1.00 \\ 
        32$\times$32 & 2727906.70 & 961.70 & 1.00 \\ 
        16$\times$32 & 965665.10 & 655.55 & 1.00 \\ 
        16$\times$16 & 475897.19 & 323.48 & 1.00 \\ 
        8$\times$8 & 97976.46 & 65.25 & 1.00 \\ 
        4$\times$4 & 24693.98 & 16.28 & 1.00 \\ \hline
    \end{tabular}
    \vspace{-4mm}
\end{table}

\subsection{Experiment customization} 
\subsubsection{Exp. Set 2: LayoutLoop Analytic Analysis}  \hfill

\textbf{Different Configurations:} LayoutLoop adopts the same architecture, dataflow constraint and mapper configurations format as TimeLoop with detailed documentations listed at \url{https://timeloop.csail.mit.edu/v4/input-formats/design}. Further, we argument LayoutLoop to support the analysis of layouts. The layout definition is shown in \secref{fig:NTLR_Layout_example_simple}. The locations of these configurations are listed below.

\squishlist
\item architecture design: ``FEATHER/LayoutLoop/configurations/arch\_designs/"
\item dataflow constraints: ``FEATHER/LayoutLoop/configurations/arch\_designs/systolic\_constraint/mapspace.yaml", the dataflow constraint needs to match hierarchies of components in the architecture design.
\item mapper: ``FEATHER/LayoutLoop/configurations/mapper/"
\item Layout: ``FEATHER/LayoutLoop/configurations/layout/"
\squishend 

\textbf{Different on-chip reordering modeling methods} are activated by enabling different global macro  
\squishlist
\item \textit{Transpose:} ENABLE\_TRANSPOSE
\item \textit{Line Rotation:} MEDUSA 
\squishend 
By default, Layoutloop assumes no on-chip reordering.

\subsubsection{Exp. Set 3: LayoutLoop Analytic Analysis} 
The provided Verilog implementation of \Accel is a parameterized scalable template, which allows users to change the shape of \Accel by modifying the input parameters at the top module  ``FEATHER/FEATHER\_RTL/RTL/feather\_top.v". Users could modify the following parameters into value from $(4,8,16,32,64)$ to investigate the area and power of \Accel at different scsales.
\begin{python}
module feather_top #(
    parameter DPE_COL_NUM  = 64,
    parameter DPE_ROW_NUM  = 64,
    ...
\end{python}




\bibliographystyle{IEEEtranS}
\bibliography{main}
\vspace{-2mm}

\end{document}